\documentclass[twocolumn]{aastex63}

\shorttitle{A duality in the origin of bulges and spheroidal galaxies}
\shortauthors{Costantin et al.}

\usepackage[]{xcolor}
\usepackage{comment}

\NewPageAfterKeywords

\begin{document}

\title{A duality in the origin of bulges and spheroidal galaxies}

\correspondingauthor{Luca Costantin}
\email{lcostantin@cab.inta-csic.es}

\author[0000-0001-6820-0015]{Luca Costantin}
\affiliation{Centro de Astrobiolog\'ia (CSIC-INTA), Ctra de Ajalvir km 4, Torrej\'on de Ardoz, 28850, Madrid, Spain}

\author[0000-0003-4528-5639]{Pablo G.~P\'erez-Gonz\'alez}
\affiliation{Centro de Astrobiolog\'ia (CSIC-INTA), Ctra de Ajalvir km 4, Torrej\'on de Ardoz, 28850, Madrid, Spain}

\author[0000-0002-8766-2597]{Jairo M\'endez-Abreu}
\affiliation{Instituto de Astrof\'isica de Canarias, 38200, La Laguna, Tenerife, Spain}
\affiliation{Departamento de Astrof\'isica, Universidad de La Laguna, 38205, La Laguna, Tenerife, Spain}
\affiliation{Dpto.~de F\'isica y del Cosmos, Campus de Fuentenueva, Edificio Mecenas, Universidad de Granada, 18071, Granada, Spain}\affiliation{Instituto Carlos I de F\'isica Te\'orica y Computacional, Facultad de Ciencias, 18071, Granada, Spain}

\author[0000-0002-1416-8483]{Marc Huertas-Company}
\affiliation{Instituto de Astrof\'isica de Canarias, 38200, La Laguna, Tenerife, Spain}
\affiliation{Departamento de Astrof\'isica, Universidad de La Laguna, 38205, La Laguna, Tenerife, Spain}
\affiliation{LERMA, Observatoire de Paris, CNRS, PSL, Universit\'e de Paris, France}

\author[0000-0001-7399-2854]{Paola Dimauro}
\affiliation{Observat\'orio Nacional, Rua General Jos\'e Cristino, 77, S\~ao Crist\'ov\~ao, 20921-400, Rio de Janeiro, Brazil}

\author[0000-0002-4140-0428]{Bel\'en Alcalde-Pampliega}
\affiliation{European Southern Observatory (ESO), Alonso de C\'ordova 3107, Vitacura, Casilla 19001, Santiago de Chile, Chile}

\author[0000-0002-2861-9812]{Fernando Buitrago}
\affiliation{Instituto de Astrof\'isica e Ci\^encias do Espa\~co, Universidade de Lisboa, Tapada da Ajuda, 1349-018, Lisbon, Portugal}

\author[0000-0002-8680-248X]{Daniel Ceverino}
\affiliation{Universidad Autonoma de Madrid, Ciudad Universitaria de Cantoblanco, 28049, Madrid, Spain}
\affiliation{CIAFF, Facultad de Ciencias, Universidad Autonoma de Madrid, 28049 Madrid, Spain}

\author[0000-0002-3331-9590]{Emanuele Daddi}
\affiliation{AIM, CEA, CNRS, Universit\'e Paris-Saclay, Universit\'e Paris Diderot, Sorbonne Paris Cit\'e, 91191, Gif-sur-Yvette, France}

\author[0000-0002-9013-1316]{Helena Dom\'inguez-S\'anchez}
\affiliation{Institute of Space Sciences (ICE, CSIC), Campus UAB, Carrer de Magrans, 08193, Barcelona, Spain}
\affiliation{Institut d'Estudis Espacials de Catalunya (IEEC), 08034, Barcelona, Spain}

\author{N\'estor Espino-Briones}
\affiliation{Departamento de F\'isica de la Tierra y Astrof\'isica, Faultad de CC F\'isicas, Universidad Complutense de Madrid, 28040 Madrid, Spain}

\author[0000-0002-4237-5500]{Antonio Hern\'an-Caballero}
\affiliation{Centro de Estudios de F\'isica del Cosmos de Arag\'on, Plaza San Juan 1, planta 2, 44001, Teruel, Spain}

\author[0000-0002-6610-2048]{Anton M.~Koekemoer}
\affiliation{Space Telescope Science Institute, 3700 San Martin Dr., Baltimore, MD 21218, USA}

\author[0000-0002-9415-2296]{Giulia Rodighiero}
\affiliation{Department of Physics and Astronomy, University of Padova, Vicolo Osservatorio 3, 35122, Padova, Italy}

%%  Abstract of the paper 

\begin{abstract}
Studying the resolved stellar populations of the different structural components which build 
massive galaxies directly unveils their assembly history.
We aim at characterizing the stellar population properties
of a representative sample of bulges and pure spheroids in massive 
galaxies ($M_{\star}>10^{10}$~M$_{\odot}$) in the GOODS-N field.
We take advantage of the spectral and spatial information provided by SHARDS and HST data
to perform the multi-image spectro-photometrical decoupling of the galaxy light.
We derive the spectral energy distribution separately for bulges and disks 
in the redshift range $0.14<z\leq1$ with spectral resolution $R\sim50$.
Analyzing these SEDs, we find evidences of a bimodal distribution of bulge formation redshifts. 
We find that 33\% of them present old mass-weighted ages, 
implying a median formation redshift $z_{\rm{form}}={6.2}_{-1.7}^{+1.5}$.
They are relics of the early Universe embedded in disk galaxies.
A second wave, dominant in number, accounts for bulges formed at median redshift $z_{\rm{form}}={1.3}_{-0.6}^{+0.6}$.
The oldest (1$^{\rm{st}}$-wave) bulges are more compact than the youngest.
Virtually all pure spheroids (i.e., those without any disk) are coetaneous with the 2$^{\rm{nd}}$-wave bulges, 
presenting a median redshift of formation $z_{\rm{form}}={1.1}_{-0.3}^{+0.3}$.
The two waves of bulge formation are not only distinguishable in terms of stellar ages, 
but also in star formation mode. All 1$^{\rm st}$-wave bulges formed fast at $z\sim6$, 
with typical timescales around 200~Myr. 
A significant fraction of the 2$^{\rm{nd}}$-wave bulges assembled more slowly, 
with star formation timescales as long as 1~Gyr.
The results of this work suggest that the centers of massive disk-like galaxies actually 
harbor the oldest spheroids formed in the Universe.
\end{abstract}

\keywords{galaxies: bulge - galaxies: evolution - galaxies: formation - galaxies: photometry - galaxies: stellar content - galaxies: structure}  

%  BODY OF PAPER

%%%%%%%%%%%%%%%%%%%%%%%%%%%%%%%%%%%%%%%%%%%%%%%%%%%%%%%%%%%
%%%%%%%%%%%%%%%%%%%%%%%%%%%%%%%%%%%%%%%%%%%%%%%%%%%%%%%%%%%
%% SECTION 1

\section{Introduction} \label{sec:section_1}

Integrated or spatially-resolved observations are typically used to infer
the stellar content of galaxies at low-redshift, reconstructing their star formation history (SFH)
with an ``archaeological'' approach \citep{Thomas2005, Rogers2010, GonzalezDelgado2015}.
Complementarily, to trace back the formation of stars in the Universe,
it is possible to compare the stellar content of similar samples of galaxies at different redshifts
\citep[the so-called ``look-back'' approach;][]{Schiavon2006, SanchezBlazquez2009, Gallazzi2014}.
In these studies, stellar population models with different 
SFHs are usually compared to the best model either obtained from the fit of the spectral energy distribution 
(SED) or derived using key spectral features which are sensitive to fundamental physical parameters 
such as age, metallicity, or $\alpha$ enhancement \citep{Kriek2011, DominguezSanchez2016}.
Both methods converge to a common picture, which shows that the most massive galaxies form at earlier epochs 
and in short timescales in the so-called ``downsizing'' scenario 
\citep{Cowie1996, PerezGonzalez2008, Thomas2010, McDermid2015}.

One fundamental piece of evidence is the size evolution of galaxies.
Massive quenched galaxies at high redshift ($z\sim2$) are a factor up to 6 times smaller in size than their counterpart
of the same mass today \citep{Daddi2005, Trujillo2006a, Trujillo2006b, Buitrago2008, vanderWel2014}.
Similarly, at all fixed galaxy masses, the more extended galaxies are younger,
more metal poor, and less $\alpha$-enhanced compared to the more compact galaxies \citep{Scott2017}.
Morphologically, at $z\sim2$ there is a bimodal distribution of galaxies:
the Universe is composed of quiescent compact pure spheroids with no clear evidence of a disk component
and star-forming clumpy disks, which gradually transform 
from disturbed to normal disk galaxies by $z\sim1$.
By contrast, at lower redshift massive galaxies ($M > 10^{10.8} $ M$_{\odot}$) 
present almost no morphological evolution.
The abundance of bulgeless galaxies decreases with redshift until almost the entire 
population presents a significant bulge component \citep{HuertasCompany2016}.

Indeed, this global view misses a crucial piece of information: galaxies are complex systems, 
which generally consist of multiple morphological components (i.e., bulges, disks, bars, etc.). 
In the simplest scenario, disk galaxies are composed by a central bulge and an outer disk.
In particular, according to the photometric definition, 
the bulge is identified as the light excess in the central part of a galaxy,
over and above the light associated to the exponential profile of the external stellar disk.
In this work we focus not only on pure spheroids, 
but we study the stellar populations of bulges
up to redshift $z=1$ in the context of their assembly history.

Spheroids are traditionally thought to arise either from a 
violent and dissipative collapse of protogalaxies \citep{Eggen1962, Larson1976},
or accumulation and rearrangement of stars in merger events \citep{Cole2000, Hopkins2009a}.
Recently, an alternative scenario emerged, since cosmological simulations 
point towards a rapid spheroid formation while high-redshift galaxies go through 
a gas-compaction phase \citep{Dekel2014, Zolotov2015, Tacchella2016}.
Considering the high density and high gas fraction of the Universe at high redshifts,
multiple linked mechanisms (i.e., minor mergers, violent disk instabilities, clump migration, counter-rotating streams, etc.) 
seem to conspire to quickly fuel the gas into the central region of the disky galaxy 
\citep{Mihos1996, Noguchi1999, Immeli2004a, Immeli2004b, Hopkins2006, Dekel2009, Ceverino2012, 
Forbes2014, Jog2014, Renaud2014, Danovich2015, Dyda2015, Wellons2015, Bournaud2016}.
Torques in the inner regions facilitate gas inflow and inward clump migration,
compacting the stellar mass in the form of a  
``blue nugget'' \citep{Barro2013, Barro2014}, growing a massive rotating stellar spheroid
\citep{Genzel2008, Ceverino2010} or a classical bulge 
\citep{Ceverino2015}, and feeding the central black hole \citep{Bournaud2011}.

In this first paper of a series, we provide robust estimations of some key properties
of bulges and pure spheroids (i.e., spheroids without an extended stellar disk), 
such as the stellar mass and the mass-weighted age, 
discussing the implications for their assembly history across cosmic time.
Key questions still need to be addressed: 
(1) When does the spheroidal population assemble?
(2) Why are some dark matter halos evolving more rapidly 
and forming massive spheroids at earlier cosmic time?
(3) Do spheroids present different properties at different cosmic epochs?
To date, answering these questions for the separate galaxy components remains very elusive, since
spatially resolved studies at high redshift are very scarce 
because very limited data (mainly broad-band only) can be gathered with spectral information.

We investigate the mechanisms which drive the evolution of spheroids (both bulges and pure spheroids)
taking advantage of the Survey for High-z Absorption Red and Dead Sources \citep[SHARDS;][]{PerezGonzalez2013},
an ESO/GTC Large Program which provided ultra-deep ($m < 26.5$ AB mag) imaging survey 
in 25 filters covering the wavelengths range 500-950 nm.
The state-of-the-art photometric multi-filter surveys, like SHARDS, 
allow us to easily study large samples of galaxies at different redshifts
with observations typically deeper than spectroscopic ones.
Indeed, the ``snapshots'' across cosmic time of spheroids properties offer
a more continuous view of their evolution.
Thanks to its 25 medium-band filters, SHARDS data set permits 
to accurately determine the main properties of the stellar populations 
of galaxies, providing a smooth SED with a resolution $R\sim50$.
The observed photometry provides a pseudo-spectrum at each pixel on the sky, 
which allows for a two-dimensional spatial analysis treating each filter independently.
Moreover, SHARDS photometry offers accurate estimates of absorption 
and emission features (i.e., Mg$_{\rm UV}$, $D4000$, [OII], and [OIII])
which are visible within the SHARDS wavelength range at high redshift.
SHARDS data are significantly deeper than spectroscopic surveys
and grant a consistent improvement in the spectral resolution with respect to broad-band studies,
mitigating the typical degeneracies which affect the inference of individual SFHs of galaxies
\citep{HernanCaballero2013, DominguezSanchez2016}.

We propose a novel approach based on the multi-band bulge/disk 
spectro-photometric decoupling of the galaxy light to derive the
SED separately for both the bulge and disk of each galaxy.
This methodology is based on recently developed techniques
of spectro-photometric decomposition \citep{Johnston2017, Tabor2017, MendezAbreu2019a}, and 
takes advantage of the high spatial resolution images from the Hubble Space Telescope (HST) 
Advanced Camera for Surveys (ACS) and Wide Field Camera 3 (WFC3), which are used as priors
for the decoupling of the bulge and disk light in SHARDS images.
Moreover, we complement the information provided by SHARDS and HST with the one at longer wavelengths 
by means of the Canada-France-Hawaii Telescope WIRCam data. 
This allows us to constrain the stellar mass and the stellar populations 
of each individual component in the sample with unprecedented accuracy.
Thus, the spectro-photometric analysis of SHARDS data represents a significant step forward in 
reaching a more exhaustive picture of galaxy formation 
and understanding the interplay between baryons and their dark-matter hosts. 

The paper is organized as follows. The sample of galaxies is characterized in Sect.~\ref{sec:section_2}.
The spectro-photometric decoupling of the galactic components, the error analysis, and 
the retrieval of the stellar population of bulges and pure spheroids is presented in Sect.~\ref{sec:section_3}.
We present and discuss our results in Sect.~\ref{sec:section_4} and Sect.~\ref{sec:section_5}, respectively.
Our conclusions are summarized in Sect.~\ref{sec:section_6}.
Throughout the paper we assume a flat cosmology with $\Omega_{\rm m} = 0.3$,
$\Omega_{\rm \lambda} = 0.7$, and a Hubble constant $H_0 = 70$ km s$^{-1}$ Mpc$^{-1}$.

%%%%%%%%%%%%%%%%%%%%%%%%%%%%%%%%%%%%%%%%%%%%%%%%%%%%%%%%%%%
%%%%%%%%%%%%%%%%%%%%%%%%%%%%%%%%%%%%%%%%%%%%%%%%%%%%%%%%%%%
%% SECTION 2

\begin{figure*}[t!]
\centering
\includegraphics[scale=0.42, trim=0cm 0cm 0cm 0cm , clip=true]{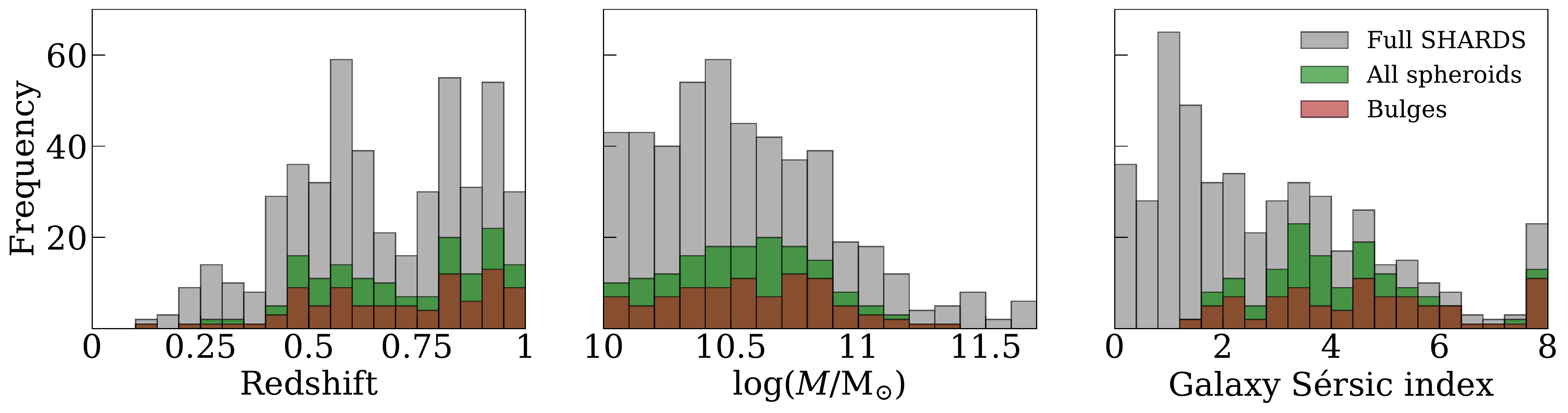}
\includegraphics[scale=0.42, trim=0cm 0cm 0cm 0cm , clip=true]{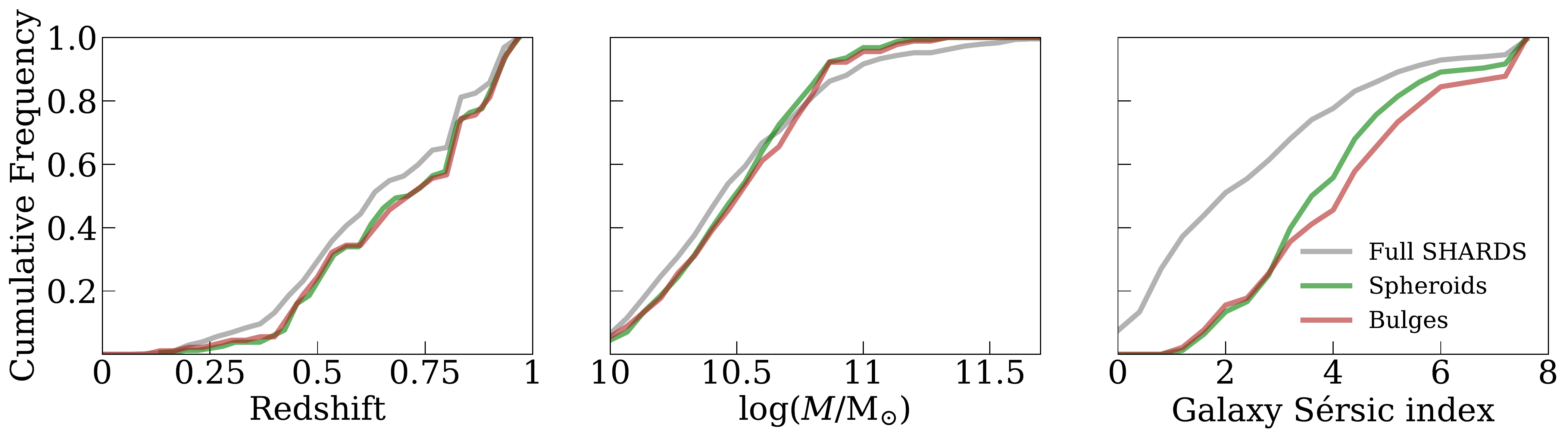}
\caption{Properties of the sample galaxies: (from left to right) galaxy redshift and stellar mass
\citep{Barro2019}, and galaxy S\'ersic index \citep{vanderWel2012}. 
(Upper panels) The gray histogram defines the mother sample of 478 galaxies in the SHARDS FoV, while
the green histogram stands for the final sample of 156 spheroids, namely, 65 pure spheroids and 
91 bulges (red histogram).
The bottom panels show the correspondent cumulative distributions.}
\label{fig:figure_1}
\end{figure*}

\section{Sample} \label{sec:section_2}

We analyze galaxies in the North field of the Great Observatory Origins Deep Survey Northern (GOODS-N)
which present photometric data provided both by HST and SHARDS.
In particular, taking advantage of the Cosmic Assembly Near-infrared Deep Extragalactic Legacy Survey
\citep[CANDELS;][]{Grogin2011,Koekemoer2011}, we use the 7 filters for HST images
(i.e., ACS F475W, F606W, F775W, and F850LP and WFC3 F105W, F125W, and F160W)
and combine them with the 25 filters for SHARDS data in the optical wavelength range 500--941 nm
\citep[see][for all details]{PerezGonzalez2013, Barro2019}.

Our mother sample consists of 478 massive ($M_{\star} > 10^{10}$ M$_{\odot}$) 
and luminous ($m_{\rm F160W} <  21.5$ mag)
galaxies at redshift $z \le1$ in the 141 arcmin$^2$ area surveyed by SHARDS \citep{Barro2019}.
In Fig.~\ref{fig:figure_1} (gray histogram) we present their mass, redshift, 
and S\'ersic index distributions \citep{vanderWel2012, Barro2019}.
We cross-match our sample galaxies with the \citet{Dimauro2018} catalogue, 
which provides photometric bulge-disk decompositions in this field-of-view, and the \citet{vanderWel2012} catalogue,
which provides photometric parameters for the galaxies.
In particular, by means of an unsupervised feature learning (deep learning) technique,
\citet{Dimauro2018} measured the best analytic model describing 
the surface brightness distribution of a given galaxy in the WFC3 F160W band:
62 are modeled with a single S\'ersic component 
(i.e., pure spheroids with bulge-over-total luminosity ratio $B/T>0.8$), 
41 galaxies which are classified as pure disks (i.e., modeled with a single exponential profile),
while 228 are modeled with two components (i.e., S\'ersic for inner bulge and exponential for outer disk).
We discard 147 galaxies, which were poorly fitted both in \citet{Dimauro2018} and \citet{vanderWel2012}.
Galaxies with very low, and thus unphysical value of their S\'ersic index, were removed from the sample.

This first paper of a series focuses on the spheroidal component, i.e., either bulges in disk galaxies or pure spheroids
(not accompanied by any other structural component),
limiting the sample selection to 290 galaxies. 
Moreover, since it is difficult to discriminate the nature of the central light prominence in nearly edge-on galaxies,
and in order to avoid the contamination from boxy-peanut structures,
we also restrict the sample of galaxies with two components to the 192 with inclination $i < 70^{\circ}$.
We further model the one-dimensional surface-brightness profile 
of galaxies with two components (see Sect.~\ref{sec:section_3-1-1}), classifying 
42 galaxies as pure disks and 15 as pure spheroids.
This criterium prevents unphysical or poor solutions derived from blind 
photometric decomposition procedure due to parameters degeneracy.
At this level, the sample is composed of  
135 two-component galaxies and 77 pure spheroids.
Finally, we visually classify these galaxies looking for signs of interaction 
and/or foreground/background contaminating objects,
removing 46 of these galaxies from our sample.

The final sample is composed of 156 galaxies, 
65 pure spheroids and 91 galaxies with a bulge and disk component,
as presented in Fig.~\ref{fig:figure_1} (green and red histograms).
Ten more galaxies were discarded from the 166 galaxies analyzed
because the 2D modeling of their individual SEDs 
provides no constraints for their stellar populations (see Sect.~\ref{sec:section_3-1-1} and~\ref{sec:section_3-2}).
The representativeness of the final sample is checked by means of a Kolmogorov-Smirnov test,
assuring that each subsample does not introduce any substantial bias in the ($z$, $M_{\star}$) parameter space
(\mbox{p-value$_z > 10\%$}, \mbox{p-value$_{M_{\star}} > 30\%$}).
Moreover, it is worth nothing that our selection clearly discards galaxies modeled with low 
values of the S\'ersic index ($n\lesssim1.5$).
This is mainly because we focus on the spheroidal component, partially due to our inclination threshold,
but also because we checked the goodness of our selection for each individual galaxy,
discarding unphysical or poorly constraint solutions.

%%%%%%%%%%%%%%%%%%%%%%%%%%%%%%%%%%%%%%%%%%%%%%%%%%%%%%%%%%%
%%%%%%%%%%%%%%%%%%%%%%%%%%%%%%%%%%%%%%%%%%%%%%%%%%%%%%%%%%%
%%% SECTION 3

\section{Analysis} \label{sec:section_3}

In this Section we present the pre-processing of the SHARDS images, the spectro-photometric
decoupling of the bulge and disk components combining HST and SHARDS information, and the analysis
of statistical errors and degeneracies.

\subsection{Spectro-photometric decoupling: bulge and disk \label{sec:section_3-1}}

For this work, we started with the v1.14.5 SHARDS images from \citet{Barro2019}. 
The reduction in all the bands redder than 800 nm was repeated (v1.15.0) 
to include larger masks for extended objects and improve the fringing 
correction and sky subtraction for galaxy outskirts. 
As a result of this, all v1.15.0 images probing wavelengths redder 
than 800~nm improved their reliability at surface brightness levels fainter than 
24.5~mag~arcsec$^{-2}$ and down to approximately 26~mag~arcsec$^{-2}$. 
This turns to be, in fact, a very important feature for our analysis of galaxy structural components.

We perform a spectro-photometric decoupling of the different galactic structures 
(i.e., bulge and disk) in the sample of massive galaxies presented in Sect.~\ref{sec:section_2} 
as a function of wavelength. Our method consists in a
robust and accurate two-dimensional photometric decomposition of the 25
medium-band SHARDS images, jointly with HST WFC3 and K-band data, 
covering the wavelength range between 400 and $2\,000~$nm. 
This analysis mimics the strategy used by the \texttt{C2D} code 
\citep{MendezAbreu2019a, MendezAbreu2019b}, where high spatial resolution 
broad-band images are used to robustly constrain spectroscopic data with lower spatial resolution. 
At the core of \texttt{C2D}, the \texttt{GASP2D} algorithm 
\citep{MendezAbreu2008, MendezAbreu2014} performs the actual photometric decompositions.

\begin{figure*}[t!]
\centering
\includegraphics[scale=0.6, trim=0cm 0cm 0cm 0cm , clip=true]{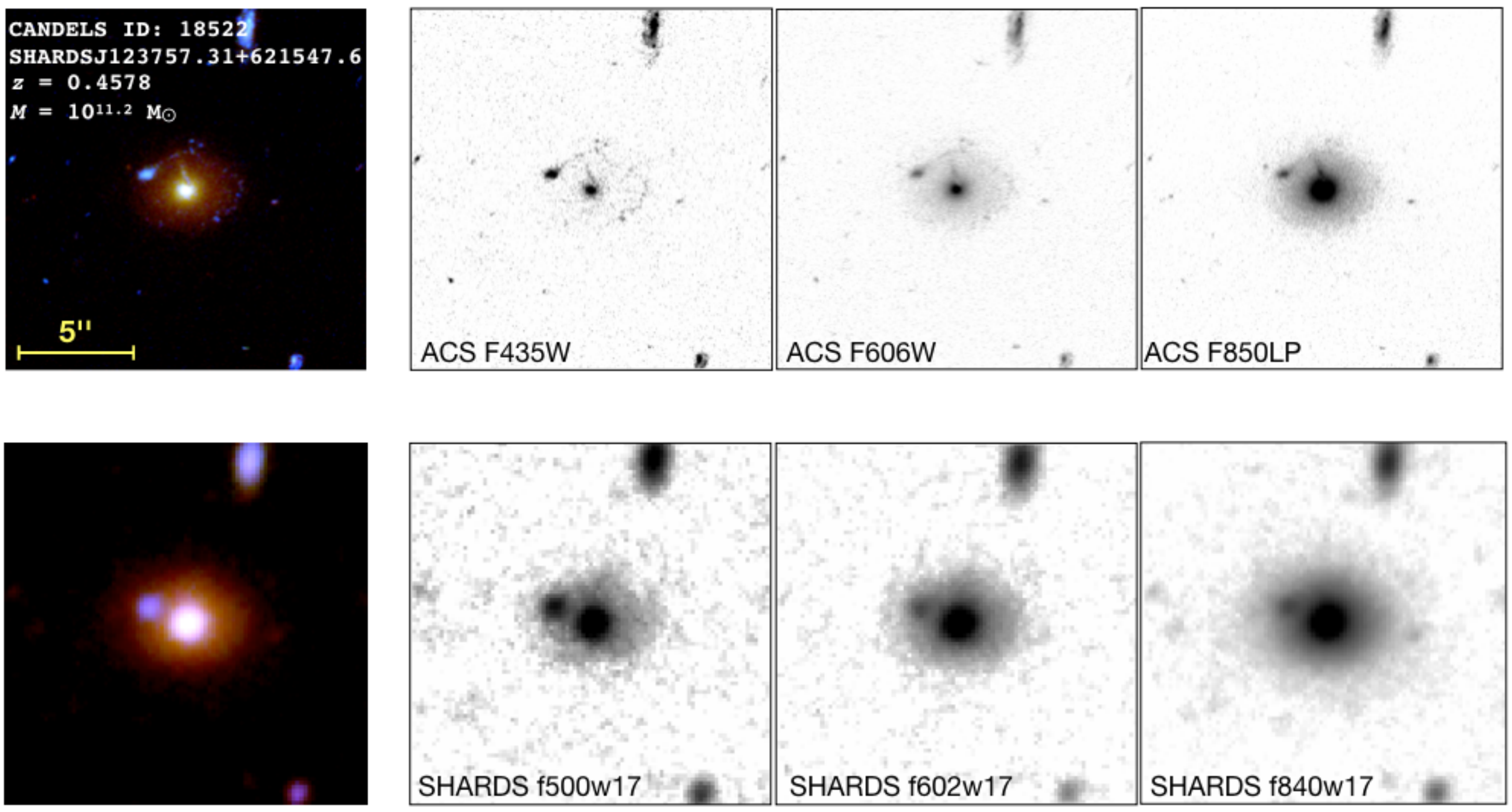}
\caption{(From left to right) Postage RGB images for a typical galaxy in our sample (GDN~18522)
and corresponding images in the three different filters: HST ACS F435W, F600W, and F850LP
(upper panels) and SHARDS f500w17, f602w17, and f840w17 (lower panels).
The galaxy is oriented North up East left and the FoV is 18$\times$18 arcsec$^2$.
\label{fig:figure_2}}
\end{figure*}

\begin{figure*}[t!]
\centering
\includegraphics[scale=0.62, trim=0cm 11cm 1cm 2.5cm, clip=true]{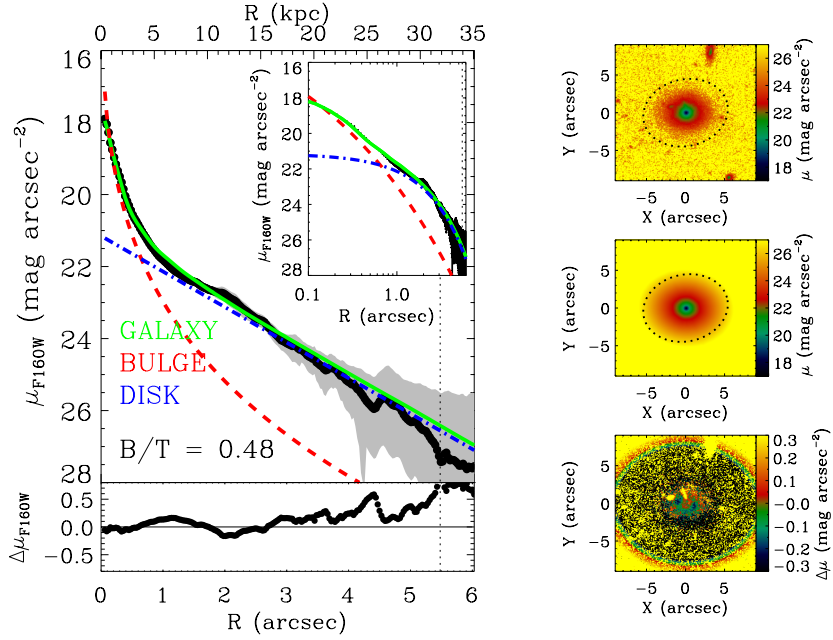}
\caption{Two-dimensional photometric decomposition of the galaxy GDN~18522 in the WFC3 F160W band
as obtained from \texttt{GASP2D}. 
The left panel shows the ellipse-averaged radial profile of the surface brightness 
measured in the observed (black dots with gray error bars) 
and PSF-convolved modeled image (green solid line) and their corresponding difference. 
The surface-brightness radial profiles of the best-fitting bulge (red dashed line) and disk 
(blue dashed-dotted line) are also shown in both linear and logarithmic scale for the distance to the center of the galaxy.
The right panels (from top to bottom) show the map of the observed, 
modeled, and residual (observed$-$modeled) surface-brightness distributions. 
The field of view is oriented with North up and East left.
The vertical black dotted lines in the left panels and the black dotted ellipses in the right panels
mark the radius where the galaxy surface-brightness reaches 26.5 mag arcsec$^{-2}$.
\label{fig:figure_3}}
\end{figure*}

\begin{figure*}[t!]
\centering
\includegraphics[scale=0.49, trim=0.5cm 12.5cm 9cm 2.5cm, clip=true]{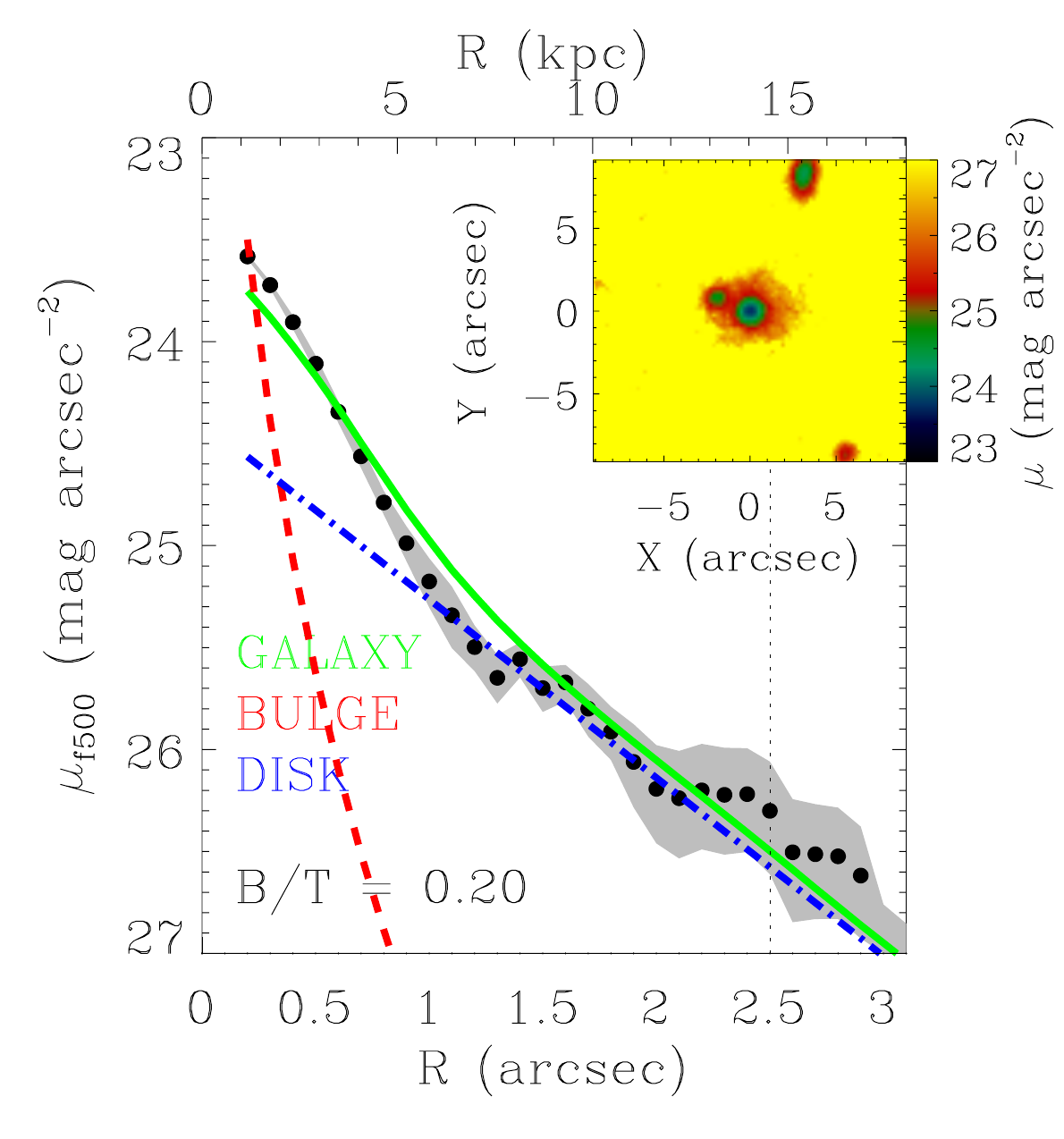}
\includegraphics[scale=0.49, trim=0.5cm 12.5cm 9cm 2.5cm, clip=true]{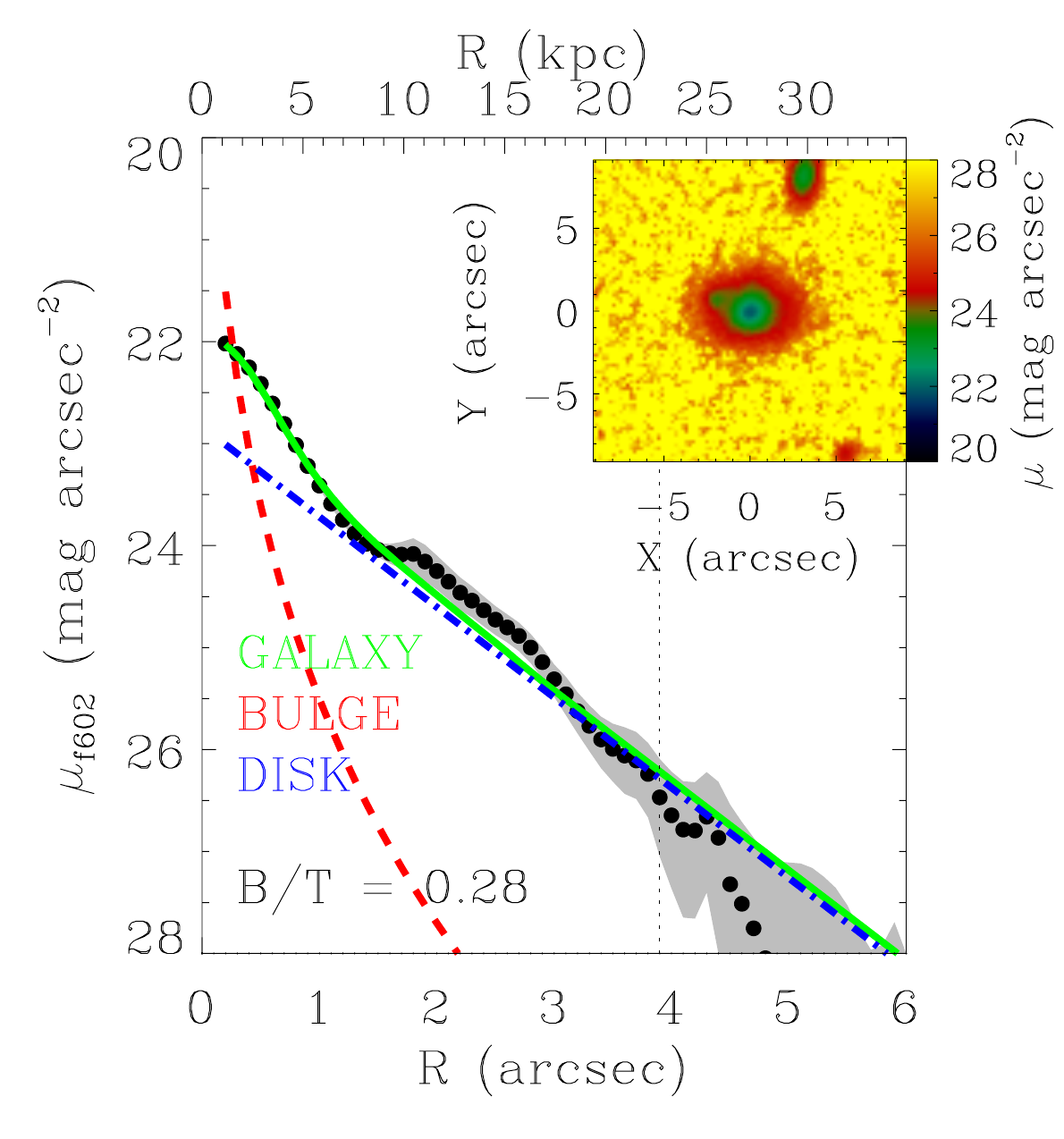}
\includegraphics[scale=0.49, trim=0.5cm 12.5cm 9cm 2.5cm, clip=true]{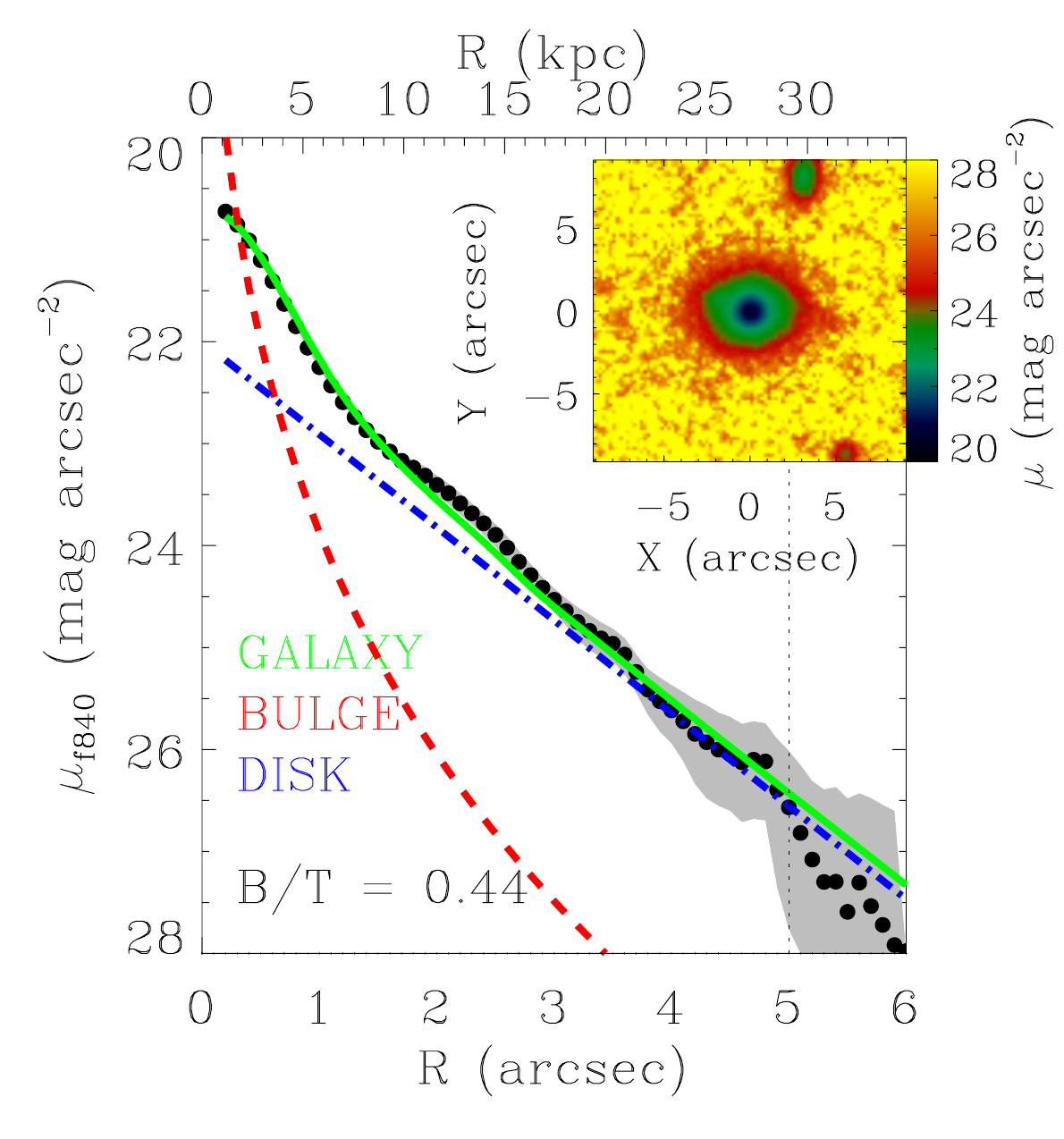}
\caption{As in Fig.~\ref{fig:figure_3}, but for SHARDS f500w17, f602w17, and f840w17 bands, respectively.}
\label{fig:figure_4}
\end{figure*}

The high flexibility of \texttt{GASP2D} allows us to represent the different galactic structures
with a variety of analytical functions.
We model the light of the bulge component with a S\'ersic function \citep{Sersic1968}:
\begin{equation}
I_{\rm bulge}(r_{\rm b}) = I_{\rm e} 10^{-b_n \left[ \left( \frac{r_{\rm b}}{R_{\rm e}} \right)^{1/n} -1 \right]} \, ,
\end{equation}
where $r_{\rm b}$ is the radius measured in the reference system of the bulge \citep[see][]{Costantin2017},
$R_{\rm e}$ is the effective radius, $I_{\rm e}$ is the intensity
at the effective radius, $n$ is the S\'ersic index describing the curvature of the profile,
and $b_n \simeq  0.868 n - 0.142$ \citep{Caon1993}.
The light of the disk component is parametrized by a single exponential \citep{Freeman1970}:
\begin{equation}
I_{\rm disk}(r_{\rm d}) = I_{\rm 0} e^{-r_{\rm d}/h} \, ,
\end{equation}
where $r_{\rm d}$ is the radius measured in the reference system of the disk \citep[see][]{Costantin2017}, while
$I_0$ and $h$ represent the central intensity and scale-length of the disk, respectively.
Both the bulge and the disk are assumed to have elliptical isophotes
centered on the galaxy center ($x_0,y_0$), with constant position angle 
$PA$ (counter-clock wise measured from the North) and constant axial ratio $q$, 
defined as the minor axis divided by major axis.

Ideally, we would like to apply a standard two-dimensional structural decomposition to each SHARDS image. 
This would imply fitting a total of 11 free parameters for each image: five for the bulge 
($I_{\rm e}$, $R_{\rm e}$, $n$, $q_{\rm b}$, $PA_{\rm b}$),
four for the disk ($I_{\rm 0}$, $h$, $q_{\rm d}$, $PA_{\rm d}$), and the galaxy center ($x_0, y_0$).
Given that the spatial resolution of the seeing-limited SHARDS images 
is around 0.9 arcsec, and considering that we are targeting galaxies at high redshift,
the degeneracies involved in determining those 11 parameters 
independently for each SHARDS bands are quite significant.
To account for this, we take advantage of the high spatial information
of multiple optical and near-infrared HST images
to derive the structural parameters of the 
bulge (i.e., $R_{\rm e}$, $n$, $q_{\rm b}$, $PA_{\rm b}$) 
and disk (i.e., $h$, $q_{\rm d}$, $PA_{\rm d}$) in our sample galaxies.
Then, we allow only the corresponding intensity to vary
in the SHARDS filters \citep[see][for a similar application using a combination of CALIFA and 
SDSS data]{MendezAbreu2019a, MendezAbreu2019b}.

\subsubsection{Synergy between HST \& SHARDS data} \label{sec:section_3-1-1}

The key strength of this work consists in combining HST and SHARDS data.
Indeed, we take full advantage of the spectral information 
provided by SHARDS medium-band images and the spatial resolution given by HST.
This allows us to retrieve individual SEDs of bulges and disks with spectral resolution $R\sim50$
up to redshift $z=1$ and down to $\log(M_{\rm gal}/$M$_{\odot}) = 10$.
In Fig.~\ref{fig:figure_2} we show an example of the different spatial information provided
by SHARDS and HST for the galaxy GDN~18522, where it is possible to appreciate the differences 
of the bulge and disk component as a function of wavelength. 

\begin{figure}[t!]
\centering
\includegraphics[scale=0.26, trim=1cm 0cm 1cm 0cm , clip=true]{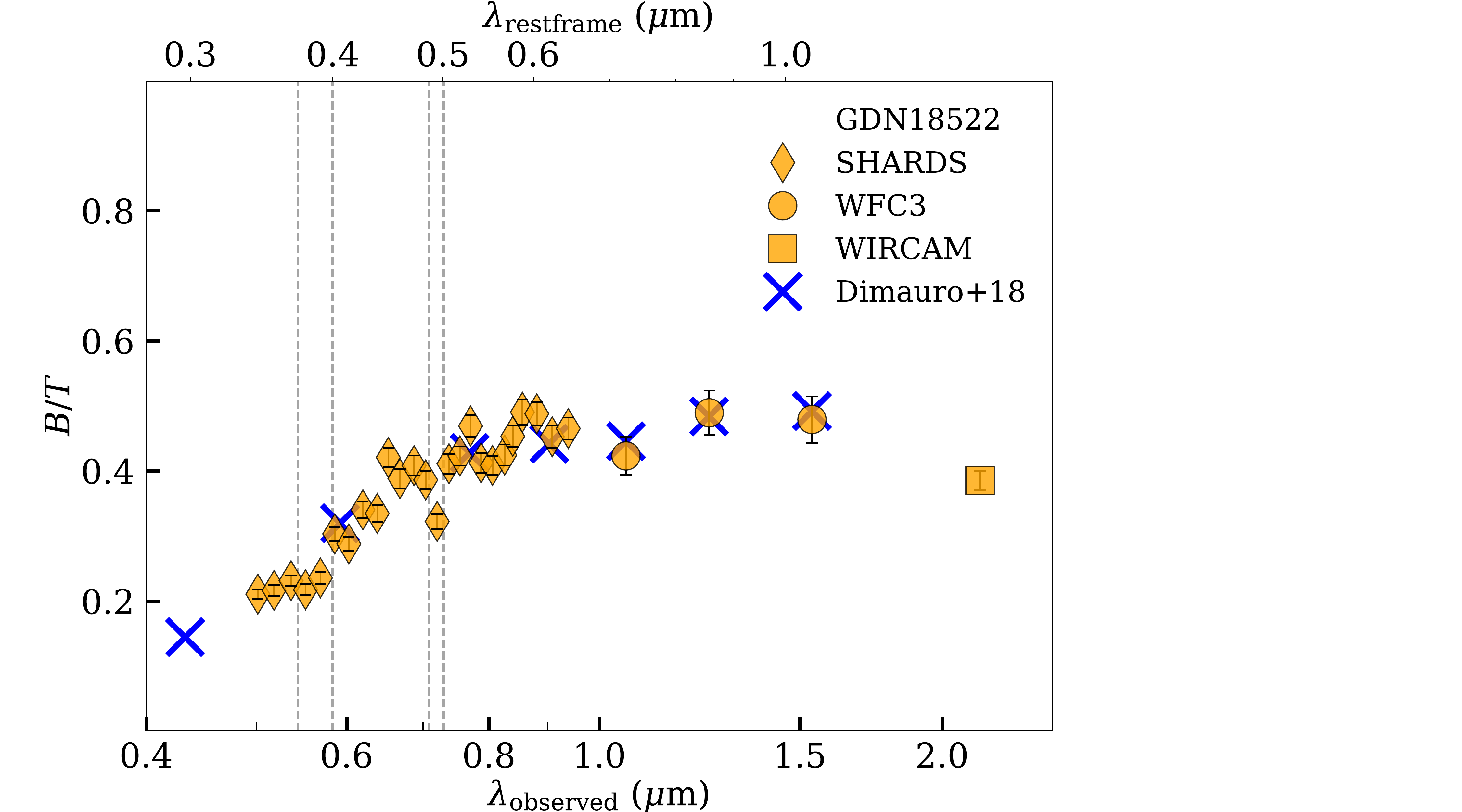}
\caption{Bulge-over-total luminosity ratio as a function of wavelength for the galaxy GDN~18522.  
Diamonds, dots, and squares represent values for SHARDS, WFC3, and WIRCam bands, respectively.
Errors are reported as 16th-84th percentile interval.
Blue crosses stand for values derived in \citet{Dimauro2018}.
From left to right, vertical dashed lines represent the location of [OII], $D4000$, H$\beta$, and [OIII] features, respectively.
\label{fig:figure_5}}
\end{figure}

As a first step in our method, we retrieve initial guesses for the bulge and disk structural parameters 
from the analysis of HST data presented in \citet{Dimauro2018}. 
We used these initial guesses as priors to characterize the bulge and disk parameter space
by means of Bayesian inference (see Sect.~\ref{sec:section_3-1-2}, for a full description).
The most probable value of the bulge and disk structural parameters (i.e., $R_{\rm e}$, $n$, $h$) 
are linearly interpolated over wavelengths, transferring the HST information to the SHARDS images.
Then, the galaxy image in each SHARDS and HST WFC3 filters is fitted using \texttt{GASP2D}. 
The intensity of the two components varies independently at each wavelength, 
while their structure is kept frozen (to the values obtained from the interpolation in the HST results). 
This is possible thanks to the exquisite 
astrometric calibration and distortion correction of the SHARDS images.
In order to better constrain the stellar mass of each individual component, we extend
the analysis to the K band using WIRCam data \citep{Hsu2019}.
For the K-band images, we use as an initial guess the structural parameters determined for the WFC3 F160W band. 
Even if this is an extrapolation of the bulge and disk properties, 
we expect mild wavelength variation of the structural parameters for our intermediate-redshift galaxy sample. 

As a result, we compute a pseudo datacube with all the spatial and spectral information 
for each component included in the fit,
as well as a datacube with the galaxy model and the residuals at each wavelength.
As an example, in Fig.~\ref{fig:figure_3} we show the radial surface-brightness profile,
the two-dimensional model, and the residual map of GDN~18522 in the WFC3 F160W band.
For comparison purposes, in Fig.~\ref{fig:figure_4} we present the radial surface-brightness profiles
obtained in three different SHARDS bands (see Appendix~\ref{sec:appendix_A} for more examples).
\citet{Dimauro2018} classified the galaxy as a two component system (probability of 0.88)
fitted with an exponential disk and a S\'ersic bulge profile
with light-weighted bulge-over-total $B/T_{F160W}=0.49$.
Accordingly, \citet{HuertasCompany2015a} classified the galaxy as spheroidal-like, with 
a probability of having a spheroid equal to 0.99 and probability of having a disk equal to 0.18.
The galaxy was flagged as a single component ``bad fit'' in \citet{vanderWel2012},
reinforcing the need of modeling the galaxy with more components to properly reproduce its structure.

The datacubes of the separated components allow us to calculate the relative contribution
on the light coming from the bulge compared to the disk in each galaxy.
Using our bulge and disk model datacube, we integrate the light of each component
up to a radius where the galaxy surface-brightness reaches 26.5 mag arcsec$^{-2}$ in the WFC3 F160W filter,
in order to maximize signal-to-noise ratio and avoid problems linked to the extrapolation of the model light.
In Fig.~\ref{fig:figure_5} we show an example of $B/T$ for GDN~18552.
Firstly, in Fig.~\ref{fig:figure_5} the spectral information provided by SHARDS medium-band filters could 
be appreciated in the distribution of $B/T$ through wavelength.
Secondly, it is remarkable how our analysis is sensitive to spectral features,
displayed as an abrupt change in the $B/T$ value in correspondence of the $D4000$ break
at 400 nm or the [OIII] emission line at 500.7 nm. 
For instance, the bulge is dimmer than expected based on a 
smooth wavelength interpolation exactly where an emission line can be present
(most probably, revealing significant star formation in the disk).
Finally, we find an overall good agreement between our values obtained from SHARDS data
and the values estimated using broad-band HST images presented in \citet{Dimauro2018}.
A quantitative test of the robustness of our decomposition 
of SHARDS images can be carried out by calculating $B/T$ ratios 
for SHARDS bands and comparing them with those obtained for HST images at 
similar wavelengths (see Fig.~\ref{fig:appendix_A0} in Appendix~\ref{sec:appendix_A}).
Considering all the galaxies in the sample, the average relative difference in the mass-weighted $B/T$
measured from the two data sets is $<16\%$ and totally consistent with the statistical errors 
(see Tables~\ref{tab:appendix_A1-1} and \ref{tab:appendix_A1-2} in Appendix~\ref{sec:appendix_A}).
The color dependence of the $B/T$ ratio observed in the HST data is also highly 
consistent with that obtained with the SHARDS bands.

\subsubsection{Error analysis} \label{sec:section_3-1-2}

It is well known that the minimization algorithms implemented in available routines for 
photometric decomposition do not usually provide a comprehensive
representation of the real errors \citep{Haussler2007, MendezAbreu2017, Costantin2017}.
Thus, for each galaxy in our sample and each SHARDS band, we build a set of mock galaxies
which take care of mimicking two sources of errors:
(a) each image is perturbed pixel by pixel according to the background noise and
(b) mock galaxies are simulated varying their structural properties
starting from the best fitted value in the WFC3 F160W band provided in \citet{Dimauro2018}.
It is worth noting that the definition of the $n$-dimensional parameter space for each galaxy results critical, 
since the structural parameters (S\'ersic index, size, and shape) are maintained frozen 
throughout our spectro-photometric analysis, allowing only for variations of the
relative intensity of the two components.

\begin{figure*}[t!]
\centering
\includegraphics[scale=0.4, trim=0cm 0cm 0cm 0cm , clip=true]{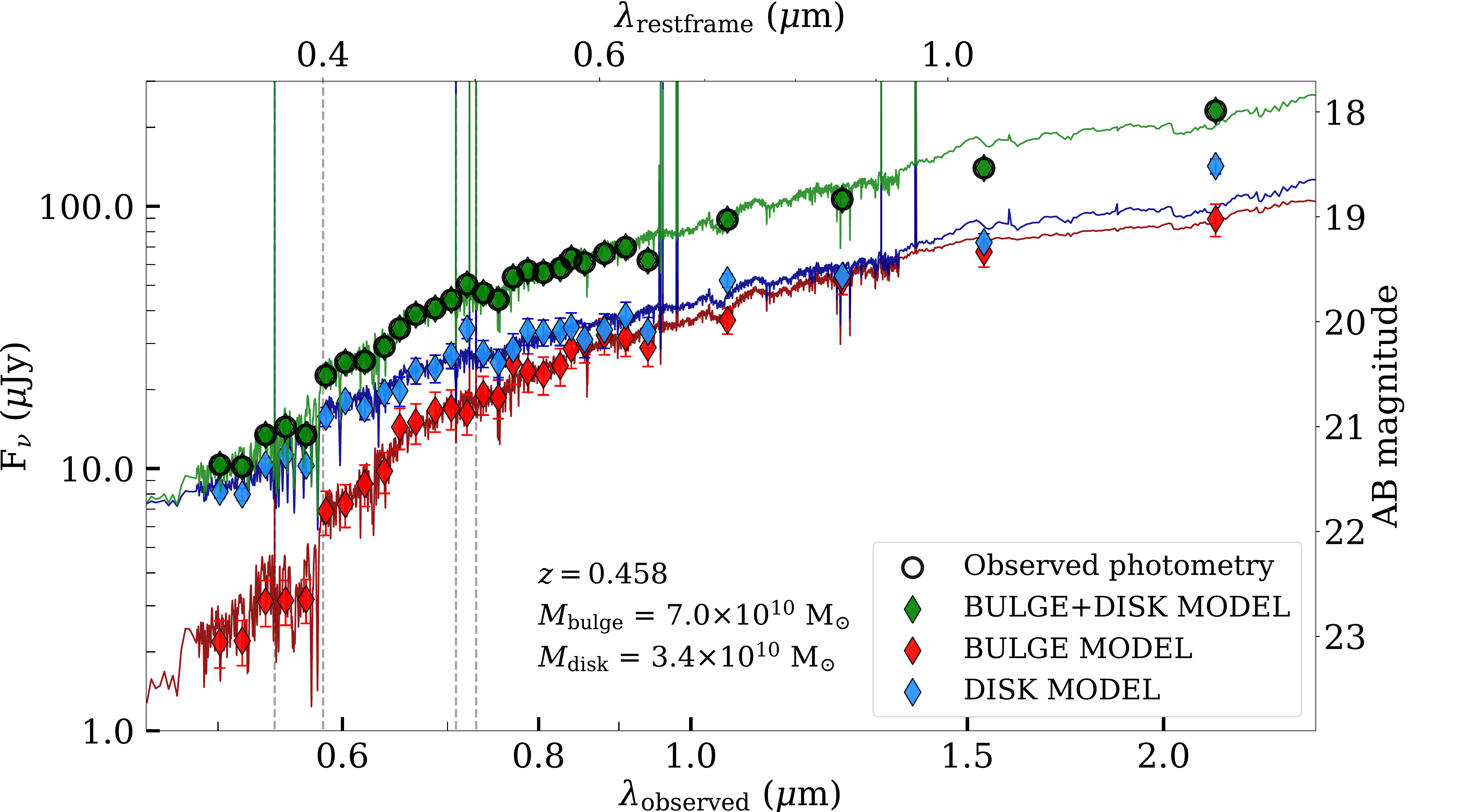}
\caption{Spectral energy distribution of the bulge (red), disk (blue), and galaxy (green) GDN~18522. 
Diamonds represent the individual photometric results of our decoupling analysis, while black circles represent 
the measured integrated photometry of the galaxy in \citet{Barro2019}.
Errors are reported as 16th-84th percentile interval.
The best model for the bulge, disk, and galaxy are shown as red, blue, and green lines.
From left to right, vertical gray dashed lines represent the location of [OII], $D4000$, H$\beta$, and [OIII] features, respectively.
\label{fig:figure_6}}
\end{figure*}

We use the No-U-Turn Sampler \citep[NUTS;][]{Hoffman2011} to find the posterior distribution 
of the model best fitting a set of data\footnote{This analysis was done using \texttt{PyMC3},
a Python open source probabilistic programming framework \citep{Salvatier2016}.}.
Given an analytic model, this Markov Chain Monte Carlo (MCMC) algorithm, 
which closely resembles a Hamiltonian Monte Carlo (HMC) method, allows us to estimate
the unknown posterior probability distribution.
For each galaxy, our model 
\begin{equation}
I_{\rm model} = I_{\rm bulge} + I_{\rm disk}
\end{equation}
is built to represent the one-dimensional surface-brightness radial profile.
In this case, the intensity of the bulge component depends on the  
effective radius, the intensity measured at the effective radius, and the S\'ersic index parameter, 
i.e., $I_{\rm bulge}$($I_{\rm e}, R_{\rm e}, n$); on the other hand, the intensity of the disk component
depends on the central intensity and the disk scale length, i.e., $I_{\rm disk}$($I_0, h$).
In details, a normal prior probability distribution is assumed for each variable in the model,
with standard deviation derived in \citet{Dimauro2018}, as well as a normal likelihood
is assumed for the total surface brightness $I_{\rm model}$.
We sample two chains for $3\,000$ tune and with $1\,000$ draw iterations.
We evaluate the MCMC convergence by means of the Gelman-Rubin diagnostic $\hat{R}$ \citep{Gelman1992},
imposing a strict criterium for convergence $\hat{R} < 1.01$.

As a consequence of the Bayes' theorem, the posterior probability distribution is evaluated,
and the probability distributions of each parameter in the model are retrieved as the end result of MCMC.
For each galaxy, we sample from this probability distribution to generate mock galaxies
with physical properties that took into account correlations between parameters and intrinsic degeneracies.
Each mock galaxy is perturbed pixel by pixel according to the background noise
and fitted using \texttt{GASP2D} as described in Sect.~\ref{sec:section_3-1-1},
using these new guesses for the structural parameters of the bulge and disk component.
We perform 100 Monte Carlo (MC) realizations for each galaxy and each SHARDS band.
This allows us to compute statistical errors for the flux of the two independent components,
taking into account degeneracies and correlations between parameters,
as well as possible biases in the photometric decomposition procedure.

The final values for the flux (and the structural parameters) of the bulge and the disk 
are retrieved as the median value of the 100 MC realizations, while the statistical errors are computed
as half of the 16th-84th percentile range, which corresponds to the standard deviation for
normally distributed errors. In the final SED, for each component and each SHARDS band,
we discard all the values of the flux which are compatible with zero according to their errors at $1\sigma$ level.
This is mostly the case of shorter wavelengths, where the signal-to-noise ratio of the images does not allow us to
consistently measure the flux, even for the total galaxy in the HST ACS F435W filter.

\begin{deluxetable*}{cccccccc}
\tablecaption{Best parameters for the sample of bulges and pure spheroids. \label{tab:table_1}}
\tablehead{
\colhead{ID} & \colhead{$\log(M_{\star}$)} & \colhead{$\bar{t}_{M}$} & 
\colhead{$z_{\rm form}$} & \colhead{$\tau$} & \colhead{$R_{\rm e}$} & \colhead{$\log(\Sigma_{1.5})$} & \colhead{type}\\
\colhead{} & \colhead{(M$_{\odot}$)} & \colhead{(Gyr)} & 
\colhead{} &\colhead{(Myr)} & \colhead{(kpc)} & \colhead{(M$_{\odot}$ kpc$^{-1.5}$)} 
}
\decimalcolnumbers
\startdata
750 & $10.6_{-0.1}^{+0.1}$ & $2.5_{-0.4}^{+0.3}$ & $1.8_{-0.2}^{+0.2}$ & $193_{-17}^{+31}$ & $1.0 \pm 0.1$ & $10.65 \pm 0.07$  & B \\
775 & $10.7_{-0.1}^{+0.1}$ & $1.4_{-0.2}^{+0.1}$ & $1.4_{-0.1}^{+0.1}$ & $194_{-21}^{+27}$ & $0.67 \pm 0.08$ & $10.98 \pm 0.08$ & B \\
912 & $11.0_{-0.1}^{+0.1}$ & $3.7_{-0.6}^{+0.4}$ & $1.5_{-0.2}^{+0.2}$ & $2490_{-270}^{+320}$ & $0.79 \pm 0.04$ & $11.20 \pm 0.03$ & B \\
2104 & $10.1_{-0.1}^{+0.1}$ & $1.1_{-0.1}^{+0.2}$ & $1.0_{-0.0}^{+0.1}$ & $200_{-21}^{+27}$ & $0.40 \pm 0.08$ & $10.7 \pm 0.1$ & B \\
5131 & $10.6_{-0.1}^{+0.1}$ & $5.7_{-0.7}^{+0.3}$ & $7.3_{-3.1}^{+4.2}$ & $203_{-24}^{+27}$ & $0.9 \pm 0.2$ & $10.7 \pm 0.1$  & B\\
6098 & $10.7_{-0.1}^{+0.1}$ & $2.2_{-0.3}^{+0.3}$ & $1.7_{-0.1}^{+0.2}$ & $198_{-22}^{+27}$ & $1.3 \pm 0.1$ & $10.55 \pm 0.06$ & B \\
6379 & $10.4_{-0.1}^{+0.1}$ & $1.9_{-0.3}^{+0.3}$ & $1.2_{-0.1}^{+0.1}$ & $203_{-23}^{+33}$ & $1.30 \pm 0.06$ & $10.22 \pm 0.03$  & B\\
\enddata
\tablecomments{Table \ref{tab:table_1} is published in its entirety in the machine-readable format.
 A portion is shown here for guidance regarding its form and content.
(1) CANDELS ID of the galaxy; (2) Stellar mass; (3) Mass-weighted age; (4) Redshift of formation;
(5) Timescale of exponentially declined SFH; (6) Effective radius; (7) Mass surface density;
(8) Type: B=bulge, PS=pure spheroid.}
\end{deluxetable*}

\subsection{Stellar populations} \label{sec:section_3-2}

The reliable estimation of the stellar mass for a large and representative sample of bulges and disks, 
as well as a proper characterization of their SFH, results critical to quantify the evolutionary process.
The typical degeneracies which affect the study of stellar populations in nearby and distant galaxies
could be mitigated by the use of photometry with higher spectral resolution 
than broad-band data \citep{Pacifici2013}. 
Using SHARDS data the statistical significance of the best solution improves up to 10-20\%
compared to the fit of broad-band data alone \citep{PerezGonzalez2013}
granting estimations of stellar masses with typical uncertainties around 0.2 dex
\citep{PerezGonzalez2008, Barro2011}.

Shortly, photometric data at different wavelengths are considered to be the end product of  
the galaxy SFH, assumed to be a declining delayed exponential:
\begin{equation}
SFR(t) \propto t/\tau^2 \, e^{-t/\tau} \, ,
\end{equation}
where $\tau$ runs from 200 Myr
to a roughly constant SFH ($\tau = 100$ Gyr). 
Increasing observational evidence and recent simulations justify
the choice of a SFH described by a rising followed by a declining phase,
irrespective of the specific parametrization \citep{Behroozi2013, Pacifici2016, 
LopezFernandez2018, Costantin2019}.
Using the synthesizer fitting code \citep[see][for all details]{PerezGonzalez2003, PerezGonzalez2008},
we compare the measured SEDs with the \citet{Bruzual2003} stellar population library, 
assuming a \citet{Chabrier2003} initial mass function integrated in the range $0.1 < M/M_{\odot} < 100$.

We allow the metallicity of the models to take discrete values $Z/Z_{\odot}$ = [0.4, 1, 2.5] 
(i.e., sub-solar, solar, and super-solar). 
This allows us to explore the trend between metallicity and
stellar mass, with more massive galaxies being more metal-rich 
than less massive ones \citep{Gallazzi2005, Gallazzi2014}.
The extinction is parametrized with a V-band attenuation assuming the extinction 
law of \citet{Calzetti2000}, with values ranging from 0 to 3 mag.
A minimization of the reduced $\chi^2$ maximum-likelihood estimator is used to
search for the best fitting model, allowing the stellar age to range from 1 Myr to 
the age corresponding to the age of the Universe at the galaxy redshift.

We run 500 MC simulations for each galaxy's component to estimate
the uncertainties in the stellar population parameters, also accounting for possible degeneracies in the solutions,
as comprehensively described in \citet{DominguezSanchez2016}.
Briefly, we model each photometric data point in the SED with a Gaussian distribution 
of width equal to the photometric uncertainty and randomly vary it, repeating the fit again 
sampling from all possible models. The final solution is computed from the analysis
of the (possible) different clusters in the $\tau$-age parameter space,
assigning a statistical significance to each of them according 
to the fraction of solutions belonging to a particular cluster.
By means of the multi-dimensional age-$\tau$-A$_V$-$Z$ space
we compute the best solutions and the corresponding uncertainties
as the median and the 68\% confidence interval of the most significant cluster's values, respectively.
It is worth noting that the estimations of stellar masses from different clusters 
of solutions are totally compatible within their uncertainties (with average scatter of 0.1 dex),
meaning that this parameter is robustly retrieved and not strongly affected by the degeneracies.

In this work, we individually fit the UV-to-NIR SEDs of bulges and disks. 
Moreover, since we are assuming that the SFH of the galaxy is
the sum of the bulge and disk one, we also fitted the total galaxy SED.
The new physical parameters of the galaxy are mostly consistent with the ones provided in \citet{Barro2019},
but could be considered as an improvement in terms of the complexity of the SFH of the galaxy.
While the observed and model values for the galaxy's SED are totally consistent, 
the little discrepancies in the galaxy physical parameters could arise from 
the different stellar models and parameter space.
As an example, in Fig.~\ref{fig:figure_6} we present the best model for the bulge, disk, and galaxy SED for the 
galaxy GDN~18522 (see also Appendix~\ref{sec:appendix_A}). 

%%%%%%%%%%%%%%%%%%%%%%%%%%%%%%%%%%%%%%%%%%%%%%%%%%%%%%%%%%%
%%%%%%%%%%%%%%%%%%%%%%%%%%%%%%%%%%%%%%%%%%%%%%%%%%%%%%%%%%%
%%% SECTION 4

\section{Results} \label{sec:section_4}

For the first time, we have applied a structure decomposition method to obtain the SEDs with spectral 
resolution $R\sim50$ for bulges and disks in a representative sample of massive galaxies at redshift $0.14 < z \leq 1$. 
Furthermore, the spectral resolution and depth of the SHARDS data allow us to measure 
absorption indices (such as $D4000$) which are closely correlated to stellar ages (see Fig.~\ref{fig:appendix_B0}). 
In this paper, we focus our analysis of these SEDs in the characterization of the spheroids 
(either bulges surrounded by a disk or pure naked spheroids) at the mentioned redshifts.
Indeed, we were able to properly reconstruct the SFHs of our sample of spheroids,
deriving fundamental physical quantities which constrain their stellar populations: 
the stellar mass ($M_{\star}$), the age $t_{\rm 0}$, the star formation timescale $\tau$, 
the metallicity ($Z$), and the dust attenuation ($A_{\rm V}$).
In Table~\ref{tab:table_1} we report the main properties of the sample spheroids.

\subsection{Mass-weighted ages and formation redshift} \label{sec:section_4-1}

We consistently characterize the SFH of both bulges and pure spheroids, 
pushing the stellar population analysis of individual structural components of galaxies to an unexplored redshift range.
In particular, we compute their mass-weighted age
\begin{equation}
\bar{t}_{M} = t_{0} -  \frac{\int_0^{t_{0}} SFR(t) \times t dt}{\int_0^{t_{0}} SFR(t) dt} \, ,
\end{equation}
where time runs from the start of the star formation in the galaxy onwards, 
i.e., $t_0$ Gyr before the Universe age corresponding to the galaxy redshift. 
This mass-weighted age represents a better approximation to the average age 
of the stellar population, which takes into account the extent of the star formation
and allows us to mitigate the age-$\tau$ degeneracy (see Table~\ref{tab:table_1}).

\begin{figure*}[t!]
\centering
\includegraphics[scale=0.55, trim=0cm 0cm 1cm 1cm , clip=true]{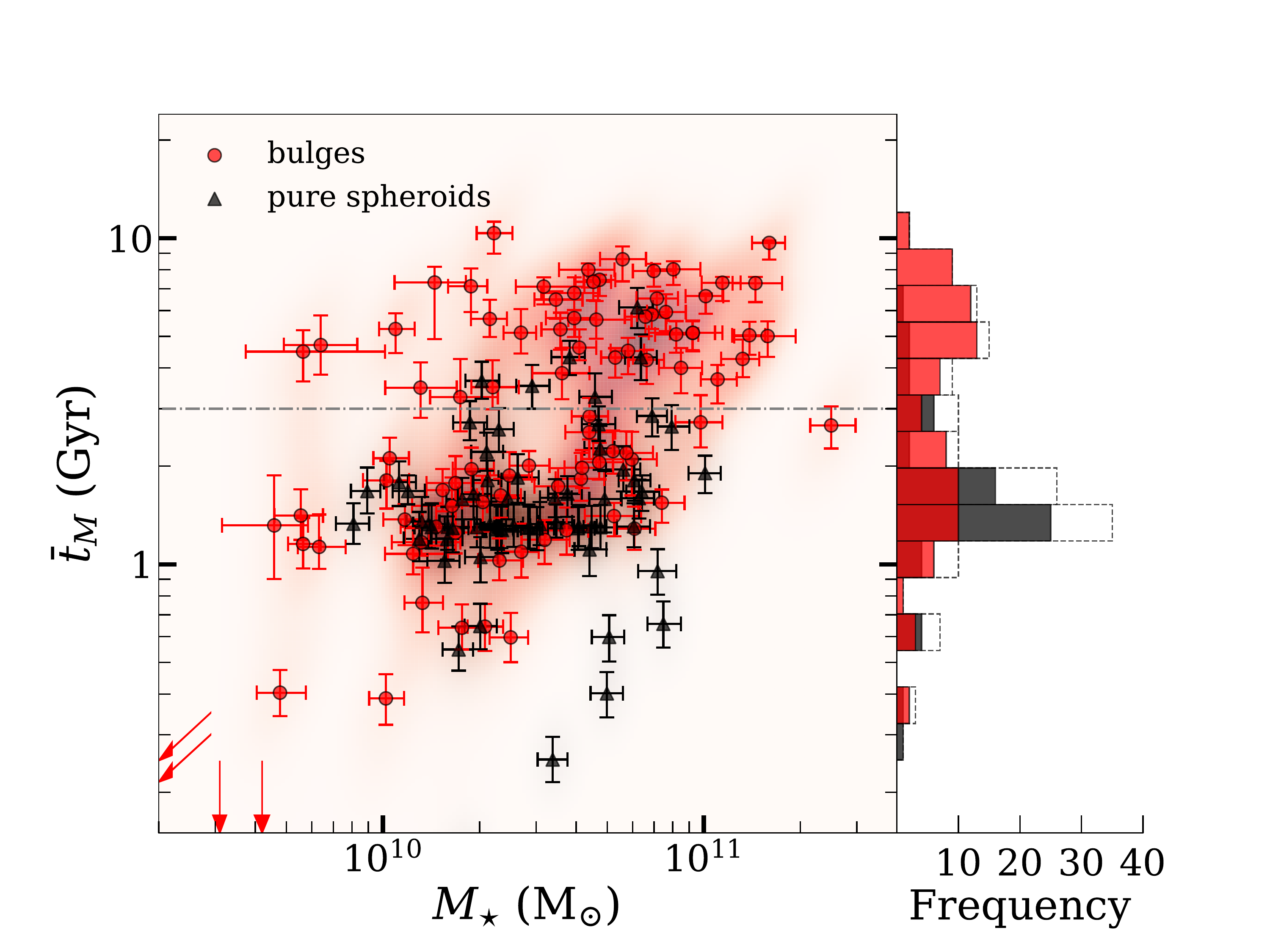}
\caption{Mass-weighted stellar ages of bulges (red dots) and pure spheroids (black triangles) 
as a function of their stellar mass. 
Errors are reported as 16th-84th percentile interval.
The red and black shaded regions show 
the distribution of the 500 MC realizations for each bulge and pure spheroid, respectively.
The gray dashed-dotted horizontal line marks $\bar{t}_{M} = 3$~Gyr. 
Red arrows mark upper limits for four systems with mass-weighted ages $\bar{t}_{M} < 100$~Myr.
The red and black histograms represent the frequency of the mass-weighted ages of
bulges and pure spheroids, respectively. The dotted black histogram shows the 
distribution for the total spheroidal population.
\label{fig:figure_7}}
\end{figure*}

%%%

In Fig.~\ref{fig:figure_7} we present the mass-weighted ages of spheroids
as a function of their stellar mass. 
The first result is that we find a very old population of bulges ($\bar{t}_{M} > 6-7$ Gyr).
In particular, bulges at redshift $0.14 < z \leq 1$ probed by our survey 
clearly displays a bimodal distribution:
52\% of them have $\bar{t}_{M} < 3$~Gyr with median value $\bar{t}_{M} = 1.8_{-0.8}^{+0.6}$~Gyr,
while 48\% have $\bar{t}_{M} > 3$~Gyr  with median value $\bar{t}_{M} = 6.3_{-1.1}^{+2.0}$~Gyr.
This behavior could reflect either a sharp difference in the formation mode of the 
two types of spheroids or be due to some kind of selection bias which we cannot identify
(see Appendix~\ref{sec:appendix_B}).
There are not massive bulges ($M_{\star} >  3 \times 10^{10}$~M$_{\odot}$) 
with young ages ($\bar{t}_{M} < 1$~Gyr).
Secondly, we find that pure spheroids are on average younger than bulges:
they have median mass-weighted ages 
$\bar{t}_{M} = 1.3_{-0.2}^{+1.2}$~Gyr and $\bar{t}_{M} = 2.7_{-1.6}^{+3.9}$~Gyr, respectively.
Considering the global spheroidal population (bulges and pure spheroids),
they have a median mass-weighted age $\bar{t}_{M} = 1.7_{-0.7}^{+3.5}$~Gyr.
When combining all spheroids, the age bimodality holds and it is dominated 
by systems less than 3~Gyr old (68\% of the entire sample):
young spheroids ($\bar{t}_{M} < 3$~Gyr) have a median mass-weighted age $\bar{t}_{M} = 1.3_{-0.6}^{+0.7}$~Gyr,
while older spheroids ($\bar{t}_{M} > 3$ Gyr) present $\bar{t}_{M} = 5.3_{-1.3}^{+2.1}$~Gyr.
At $M_{\star} \gtrsim 7 \times10^{10}$~M$_{\odot}$ the population of old spheroids dominates,
while less massive systems show similar masses when divided in old and young ones.
Four spheroids are caught in the middle of their formation process ($\bar{t}_{M} < 100$~Myr) at the redshift of observation. 

%%%%%%%%%%%%%%%%%%%%%%%%%%%%%%%%%%%%%%%%%%%%%%%%%%%%%%%%%%

\begin{figure*}[t!]
\centering
\includegraphics[scale=0.55, trim=0cm 0cm 0cm 0cm , clip=true]{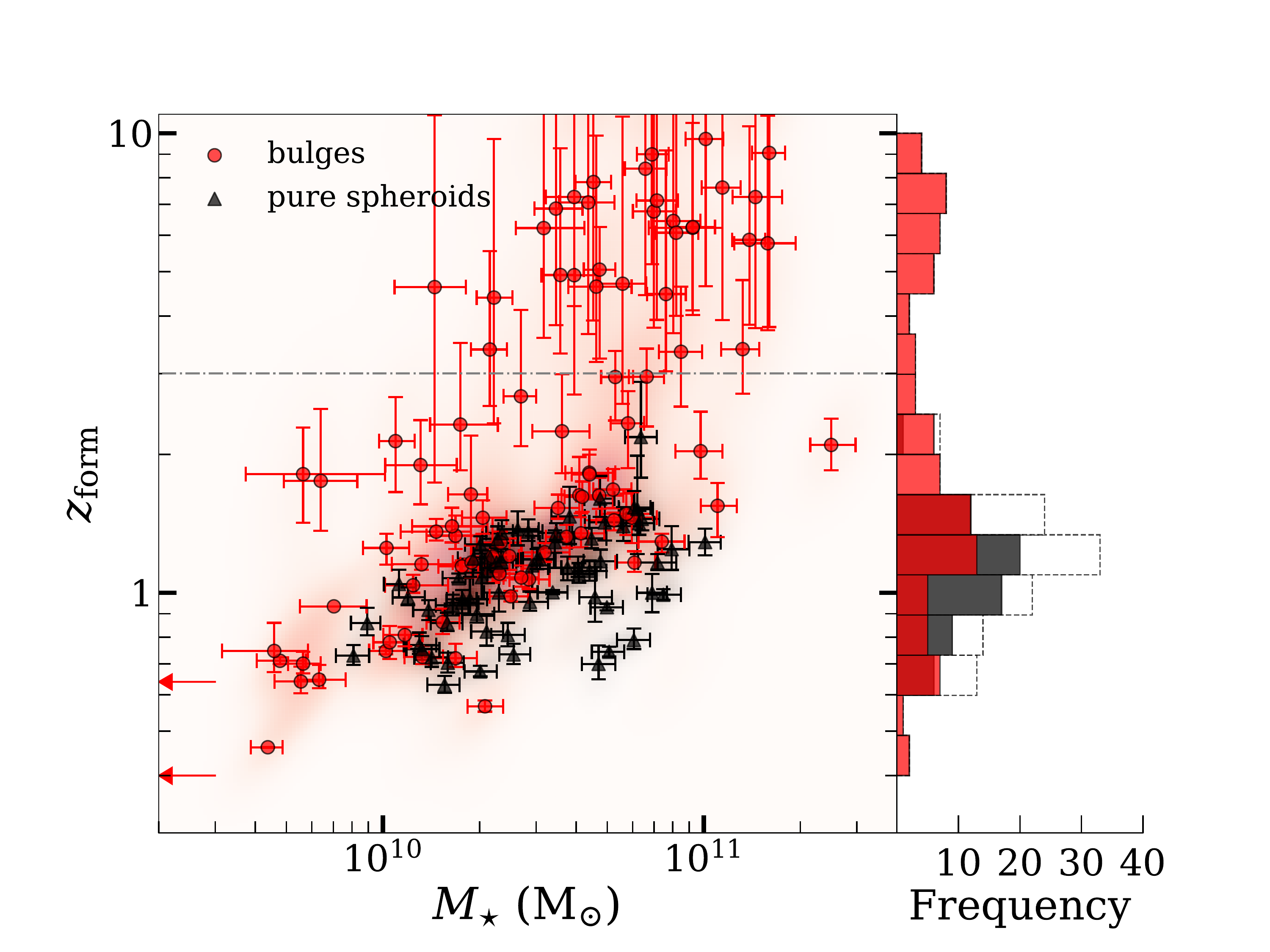}
\caption{Formation redshift of bulges (red dots) and pure spheroids (black triangles) 
as a function of their stellar mass. Errors are reported as 16th-84th percentile interval.
The red and black shaded regions show 
the distribution of the 500 MC realizations for each bulge and pure spheroid, respectively.
The gray dashed-dotted horizontal line marks
the formation redshift $z_{\rm form} = 3$.
Red arrows mark upper limits for two systems with masses $M_{\star} < 2\times10^{10}$~M$_{\odot}$.
The red and black histograms represent the frequency of the formation redshift of
bulges and pure spheroids, respectively.
The dotted black histogram shows the distribution for the total spheroidal population.
\label{fig:figure_8}}
\end{figure*}

\begin{figure*}[t!]
\centering
\includegraphics[scale=0.75,  trim=0cm 0cm 0cm 0cm, clip=true]{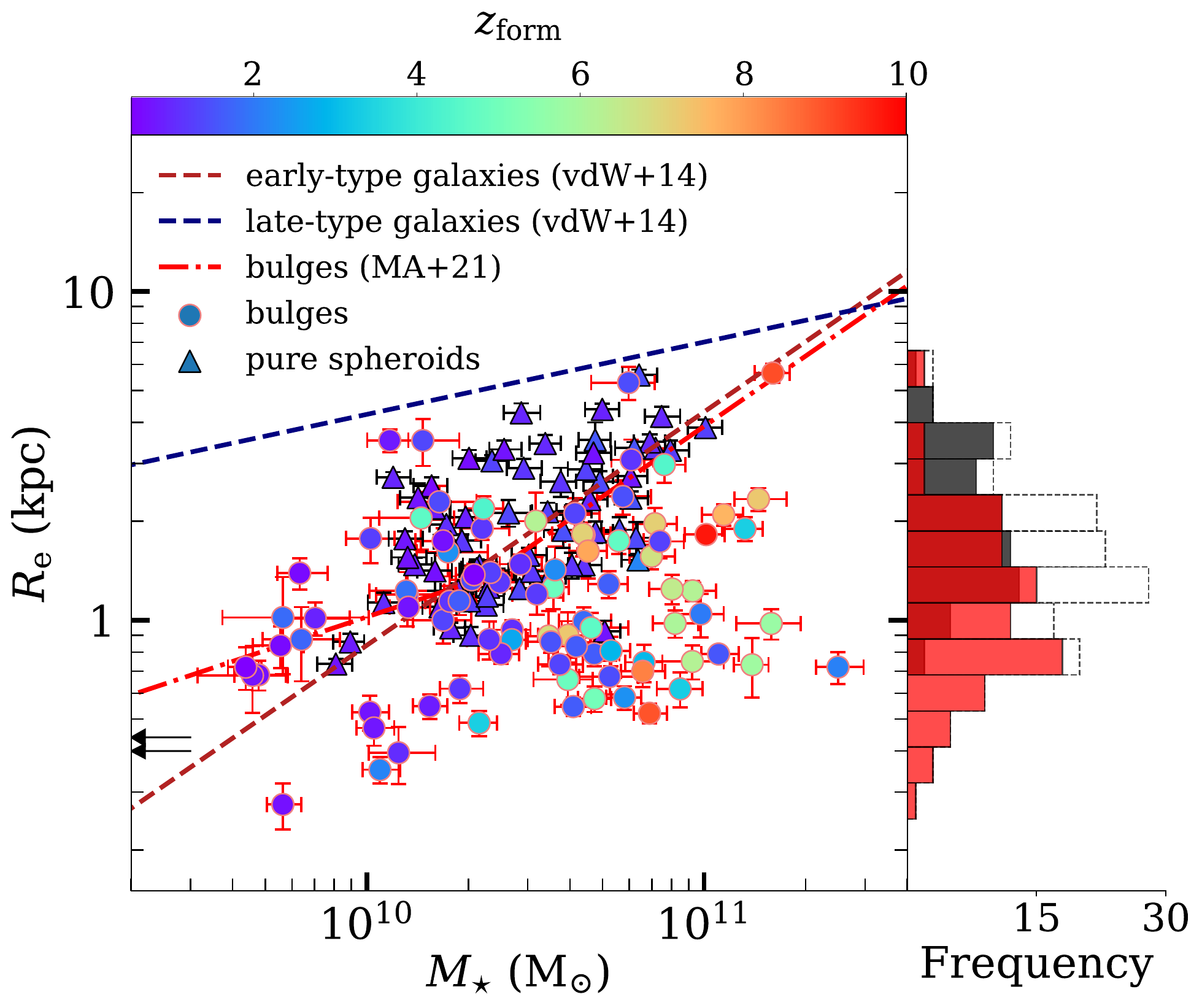}
\caption{Mass-size relations for bulges (dots) and pure spheroids (triangles), 
color-coded according to their formation redshift $z_{\rm form}$.
Errors are reported as 16th-84th percentile interval.
The dark red and blue dashed line correspond to the best-fitting trend for early-type and late-type galaxies 
at redshift $z=0.75$ in \citet{vanderWel2014}, respectively.
The light red dashed-dotted line stands for the best-fitting trend of bulges at redshift $z\sim0$ in \citet{MendezAbreu2021}.
Black arrows mark upper limits for two systems with masses $M_{\star} < 2\times10^{10}$~M$_{\odot}$.
The red and black histograms represent the frequency of the mass-weighted ages of
bulges and pure spheroids, respectively. The dotted black histogram shows the 
distribution for the total spheroidal population.
\label{fig:figure_9}}
\end{figure*}

It is worth to remember that our sample of galaxies spans along a range in redshift $0.14 < z \leq 1$ 
(see Fig.~\ref{fig:figure_1}) which translates to a wide range in cosmic time ($t \sim 6$~Gyr).
Thus, since we are dealing with galaxies observed at different epochs, 
we derive the redshift corresponding to mass-weighted ages (i.e., formation redshift $z_{\rm form}$)
to better understand their evolutionary pathways (see Table~\ref{tab:table_1}).
In Fig.~\ref{fig:figure_8} we show the correlation between formation redshift
and stellar mass of both bulges and pure spheroids.
In this way, we confirm the observation of a very old population of bulges.
The bimodality in formation redshift reflects the one in mass-weighted ages:
67\% of bulges have $z_{\rm form}<3$ with median value $z_{\rm form} = 1.3_{-0.6}^{+0.6}$,
while 33\% have $z_{\rm form}>3$ with median value $z_{\rm form} = 6.2_{-1.7}^{+1.5}$.
At $M_{\star} > 7 \times 10^{10}$ M$_{\odot}$, 67\% of bulges are formed 
at redshift $z_{\rm form} > 3.8$.
We find that pure spheroids are formed (on average) later than bulges.
The totality of pure spheroids builds up at redshift $z_{\rm form} \lesssim 2$,
with a median formation redshift  $z_{\rm form} = 1.1_{-0.3}^{+0.3}$.
Considering the global spheroidal population (bulges and pure spheroids),
19\% of the systems formed in a first wave at redshift $z_{\rm form} > 3$ 
(with median $z_{\rm form} = 6.2_{-1.8}^{+1.5}$),
while the majority of spheroids (81\%) assembles in a second wave at redshift $z_{\rm form} < 3$ 
(with median $z_{\rm form} = 1.2_{-0.4}^{+0.4}$).
Finally, there is a positive trend between formation redshift and stellar mass.
For $M_{\star} > 7 \times 10^{10}$ M$_{\odot}$ the spheroidal population is dominated by 
systems formed at $z_{\rm form}>3$ (77\%), while at lower mass 88\% of the population
is formed at $z_{\rm form}<3$.

We tested that the bimodality in the mass-weighted age and formation redshift of
our bulges is not biased by the redshift of the observed galaxies.
The two waves of bulge formation are also observed when dividing 
the sample in two redshift ranges, $z\leq0.75$ and $z>0.75$ (see Appendix~\ref{sec:appendix_B}).
On the other hand, pure spheroids do not display a bimodal distribution
and have similar mass-weighted ages and formation redshifts as the second wave of bulges.

In summary, we find an old population of bulges,
being 33\% of them already in place by $z_{\rm form}>3$.
Spheroids in the redshift range $0.14<z \leq1$ display a bimodal distribution:
systems with median mass-weighted age $\bar{t}_{M} = 1.3$~Gyr, formed at median redshift $z_{\rm form} = 1.2$, 
and considerably older bulges (``embedded'' spheroids), with a median mass-weighted age of $5.3$~Gyr, 
formed at median redshift $z_{\rm form} = 6.2$. 
Pure spheroids (which do not present a disk) belong to the first type, 
having median mass-weighted age is $\bar{t}_{M} = 1.3_{-0.2}^{+1.2}$~Gyr and being
all of them formed at $z\lesssim2$.

Hints for an early formation of bulges in disk galaxies were already provided by \citet{Morelli2016}
by direct age measurements in a sample of 12 local galaxies. Despite the low statistics,
they identify a double population of bulges: 
7 young bulges ($2-7$~Gyr) with solar metallicity and 
5 old bulges ($\gtrsim 13$~Gyr) with a large spread in metallicity. 
Nonetheless, regardless of the actual bimodality, one of the main results of this work 
is that a fraction of bulges formed at very high redshift.
Despite studying a small sample of 10 star-forming galaxies at redshift $0.45 < z < 1$, 
\citet{Mancini2019} already saw hints that quiescent bulges
present ages approaching the age of the Universe at the time of observation.

Our results agree and provide a deeper insight about the 
trend recently found by independent spectroscopic observation of galaxies 
from redshift $z=0.7$ to redshift $z = 2.5$ \citep{Gallazzi2014, Carnall2019, Belli2019}.
In particular, \citet{Gallazzi2014} studied a sample of $\sim70$ galaxies at redshift $z\sim0.7$,
finding average r-band light-weighted ages from $\sim2.3$ Gyr ($z_{\rm form} \sim 1.3$) at log($M_{\star}$/M$_{\odot}$) = 10.5
to $\sim2.9$ Gyr ($z_{\rm form} \sim 2.2$) at log($M_{\star}$/M$_{\odot}$) = 11.4.
Considering the average bulge-over-total mass ratio $(B/T)_{\rm mass} = 0.51$ of our sample galaxies,
the mass range proven by our 2$^{\rm nd}$-wave bulges overlaps with their sample galaxies.
On the other hand, our first wave of spheroid formation could be compared
with higher redshift observations. 
\citet{Carnall2019} derived the formation redshift of 75 massive (log($M_{\star}$/M$_{\odot}$) $>$ 10.3)
UVJ-selected galaxies at redshifts of $1.0 < z < 1.3$
finding a population of old systems ($z_{\rm form} > 3$) which have higher stellar masses
compared to a younger population which dominates their sample.
This trend is also consistent with the one provided by \citet{Belli2019}
studying 24 quiescent galaxies at $1.5 < z < 2.5$.
It is worth noting that the scenario holds besides the differences in the details of the SFHs
and the different spectro-photometric data sets, providing really strong constraints for cosmological simulations
aiming at reproducing the observed Universe.

\subsection{Morphological properties \label{sec:section_4-2}}

In order to understand what drives the formation redshift 
of the substructures of massive galaxies and the possible differences in the morphological properties 
of older and younger spheroids, we plot in Fig.~\ref{fig:figure_9}
their mass-size relations.
For pure spheroids we use sizes according to their half-light radius 
in the WFC3 F160W band from \citet{Dimauro2018},
while we characterize the size of the bulge using the 
best value of its half-light radius obtained in the spectro-photometric 
decoupling using the WFC3 F160W filter (see Sect.~\ref{sec:section_3-1-2}).
The stellar masses of spheroids are derived from the stellar population analysis described in Sect.~\ref{sec:section_3-2}.

\begin{figure}[t!]
\centering
\includegraphics[scale=0.29, trim=0cm 0cm 0cm 0cm , clip=true]{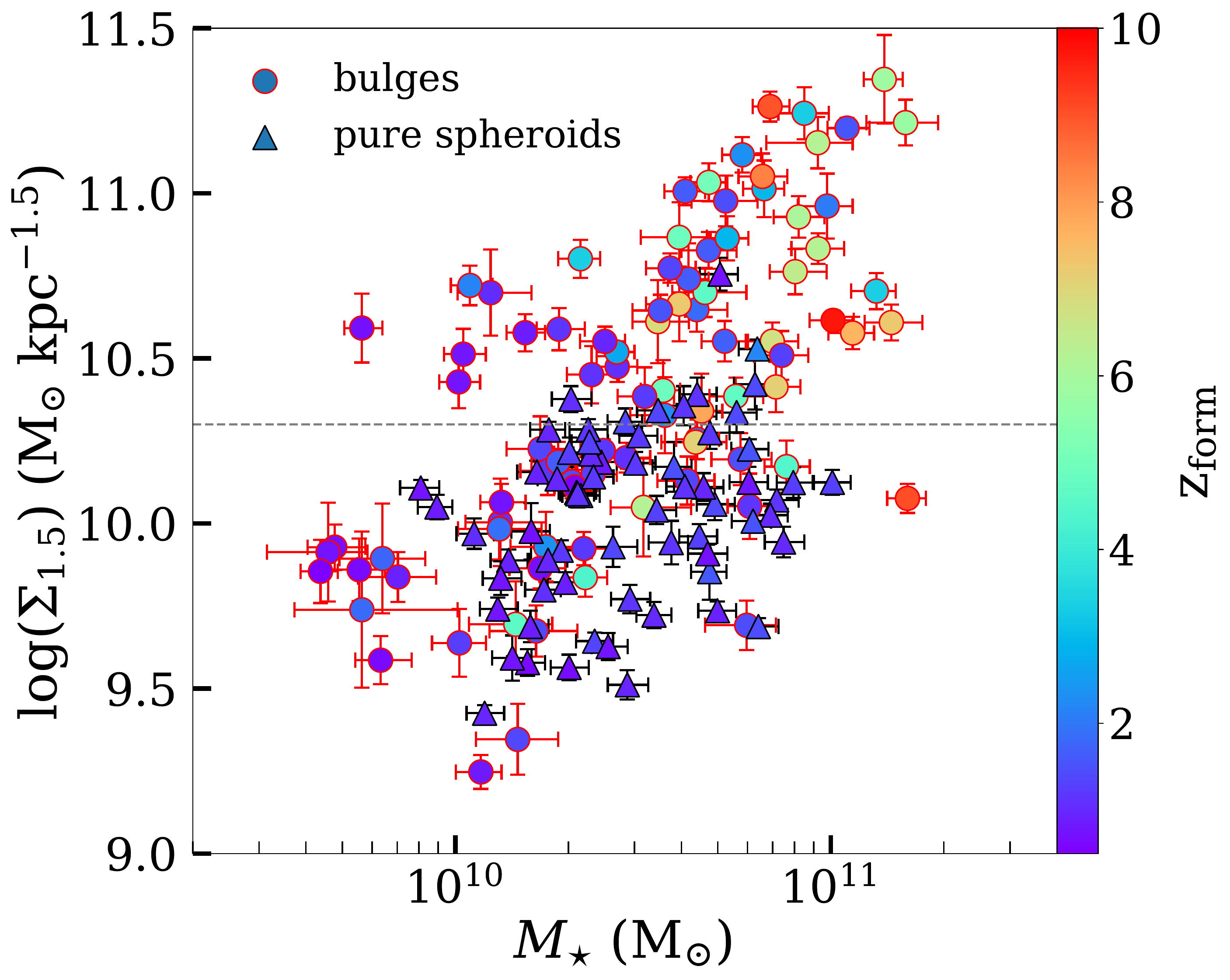}
\caption{Mass surface density $M R_{\rm e}^{- 1.5}$ of bulges (dots) and pure spheroids (triangles) 
as a function of their mass, color-coded according to their formation redshift $z_{\rm form}$.
Errors are reported as 16th-84th percentile interval.
Systems are separated between compact and extended ones 
at $\log(\Sigma_{1.5}) = 10.3$ M$_{\odot}$ kpc$^{-1.5}$ \citep[dashed gray horizontal line;][]{Barro2013}.}
\label{fig:figure_10}
\end{figure}

We compare the position of bulges and pure spheroids
in the mass-size plane with the trend expected for early and late-type galaxies described in \citet{vanderWel2014}.
The two populations actually correspond to star-forming and quiescent galaxies
classified by means of a rest-frame colors selection \citep[see][]{Wuyts2007, Williams2009}.
In particular, we use the best-fitting relation at redshift $z = 0.75$, 
but similar trends are observed when considering the evolving mass-size relationships 
described in other works \citep{Buitrago2008, Damjanov2009, vanderWel2014}.
Both bulges and pure spheroids lie in the region of early-type galaxies,
having masses and sizes which are not compatible with late-type systems.
The bulge population is more compact than its counterpart at redshift $z\sim0$, 
consistent with an increased star formation activity in the galaxy most internal regions at earlier times.
We note in Fig.~\ref{fig:figure_9} that they are offseted from the best-fitting relation in
\citet{MendezAbreu2021} toward lower sizes and higher masses.

Interestingly, at fixed stellar mass, bulges are on average smaller 
than pure spheroids in the same redshift range.
The first ones have median sizes $R_{\rm e} = 1.0_{-0.4}^{+0.9}$~kpc,
while the latter show $R_{\rm e} = 1.9_{-0.6}^{+1.4}$~kpc.
Indeed, they are two different populations in terms of size, 
as proved by means of the Kolmogorov-Smirnov statistics (p-value $< 1\%$).
Bulges extend to lower sizes, that is, 
pure spheroids are not found among the smallest spheroids:
92\% of them have $R_{\rm e} > 1$~kpc, while 48\% of 
bulges have $R_{\rm e} \leq 1$~kpc.
Although it seems that there is a continuity between these two classes,
a multivariate Wald-Wolfowitz test reinforces the result
that they are not the same population.
At low redshift, various studies confirm that elliptical galaxies and bulges are
two different populations in terms of their size \citep{Gadotti2009, MendezAbreu2021}.
Local bulges and massive elliptical galaxies seem to
follow offseted mass-size relations and occupy different loci of the Fundamental Plane \citep{Djorgovski1987},
providing clues to their different evolutionary processes \citep{Gadotti2009}.
Nevertheless, it has to be noted that this result is less strict for low-luminosity elliptical galaxies,
which fits in the picture that elliptical galaxies actually harbor a more compact (and probably older)
component in their core \citep{delaRosa2016}.

Regarding their evolution, young spheroids ($z_{\rm form} < 3$) 
have median sizes $R_{\rm e} = 1.4_{-0.6}^{+1.5}$~kpc
and older ones ($z_{\rm form} > 3$) have $R_{\rm e} = 1.3_{-0.6}^{+0.8}$~kpc.
In particular, we find a lack of extended ($R_{\rm e} > 2.5$~kpc) spheroids among 
the population formed at the highest redshift.

The most important result we can infer from the mass-size relation in Fig.~\ref{fig:figure_9}
is that 1$^{\rm st}$-wave bulges are more compact than 2$^{\rm nd}$-wave ones: 
they are not only slightly smaller, but also more massive.
Since we find no clear trend neither with the S\'ersic index 
of the galaxy nor the S\'ersic index of bulge,
to properly quantify the evolution of spheroids, we plot in Fig.~\ref{fig:figure_10} their
formation redshift as a function of their mass surface density $\Sigma_{\alpha} = M R_{\rm e}^{- \alpha}$ (see Table~\ref{tab:table_1}).
According to the definition in \citet{Barro2013}, we define  $\Sigma_{1.5}$ ($\alpha = 1.5$), 
such as it lies between the mass surface density $M/R_{\rm e}^2$ and $M/R_{\rm e}$,
both of which strongly correlated with color and SFR up to high redshifts \citep{Franx2008}.
Following \citet{Barro2013}, we could identify our spheroids with their
compact population of quiescent galaxies, defined by $\log(\Sigma_{1.5}) > 10.3$~M$_{\odot}$~kpc$^{-1.5}$.
Young bulges ($z_{\rm form} < 3$) have 
median mass surface density $\log(\Sigma_{1.5}) = 10.2_{-0.4}^{+0.5}$~M$_{\odot}$~kpc$^{-1.5}$,
while older ones have median mass surface density $\log(\Sigma_{1.5}) = 10.6_{-0.4}^{+0.4}$~M$_{\odot}$~kpc$^{-1.5}$.
While bulges span a wider range of mass surface densities (median values
$\log(\Sigma_{1.5}) = 10.4_{-0.6}^{+0.5}$~M$_{\odot}$~kpc$^{-1.5}$),
86\% of pure spheroids have $\log(\Sigma_{1.5}) < 10.3$~M$_{\odot}$~kpc$^{-1.5}$.

Summarizing, the 1$^{\rm st}$-wave bulges are more compact than the 2$^{\rm nd}$-wave bulges,
and bulges are more compact than pure spheroids.

%%%%%%%%%%%%%%%%%%%%%%%%%%%%%%%%%%%%%%%%%%%%%%%%%%%%%%%%%%%
%%%%%%%%%%%%%%%%%%%%%%%%%%%%%%%%%%%%%%%%%%%%%%%%%%%%%%%%%%%
%% SECTION 5

\begin{figure}
\centering
\includegraphics[scale=0.31]{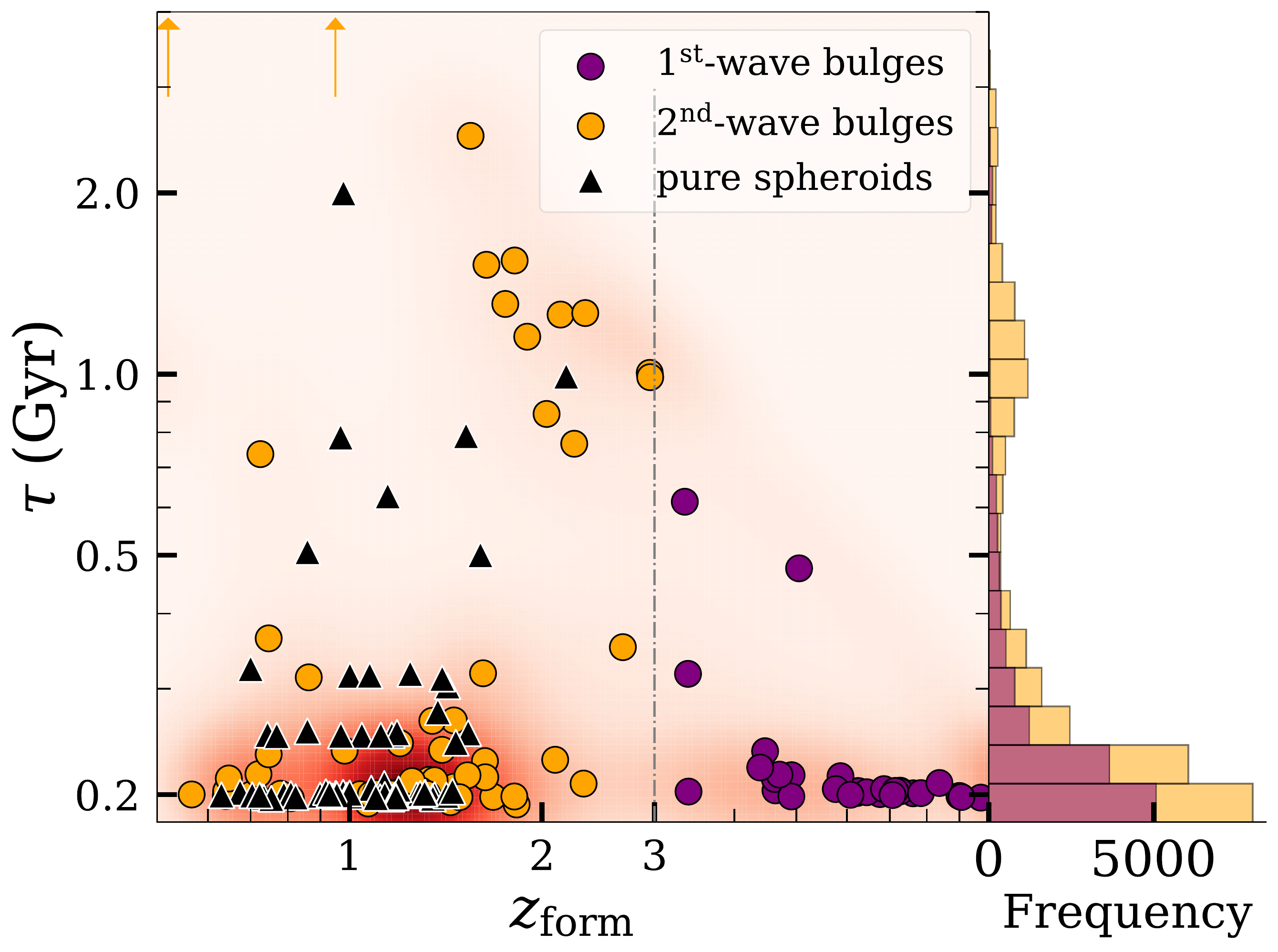}
\caption{Distribution of timescales as a function of formation redshift
for our bulges (first wave: purple dots; second wave: orange dots) and pure spheroids (black triangles).
The red shaded region shows the density distribution of the 500 MC realizations for each bulge.
The purple and orange histograms represent the frequency of the timescales
for all the MC realizations of 1$^{\rm st}$ and 2$^{\rm nd}$-wave bulges, respectively.}
\label{fig:figure_11}
\end{figure}

\section{Discussion} \label{sec:section_5}

In this first paper of a series we focus on studying independently 
the separate structures in disk galaxies up to redshift $z=1$,
which allows us to compare the properties of bulges with pure spheroids.
Albeit the plethora of high resolution data collected at redshift $z\sim0$, it still remains difficult to resolve 
SFHs when galaxies are older than 5 Gyr due to the similarity of their stellar spectra.
On the contrary, the window $0.14 < z \leq 1$ gives us the advantage of
discerning different stellar populations of galaxies up to higher formation redshift,
simply because the stellar ages are bounded by the age of the Universe at the epoch of observation.
Moreover, SHARDS spectral resolution and wavelength coverage range permit
to probe young stellar populations due to the presence of optical spectral features in their rest-frame SEDs.

\subsection{Bulges form in two waves}

In this work we prove that the population of more massive spheroids ($M_{\star} > 7 \times 10^{10}$~M$_{\odot}$) 
is mainly composed by systems already in place at very early cosmic time ($z_{\rm form} \gtrsim 4$), 
while the low-mass end is dominated by spheroids formed in a prolonged timespan of $\sim8$~Gyr.
Our result suggests that in the early phase of galaxy evolution the conditions were very favorable for rapidly 
assemble and quench the building blocks of today's galaxies.
Indeed, the population of bulges formed at $z_{\rm form} > 4$
can be considered as the relic of the early Universe.

Star formation can be very rapidly quenched at earlier cosmic time,
as suggested by the presence of quiescent galaxies at redshifts $z\gtrsim4$ \citep{Straatman2014}.
They are typically highly-compact elliptical-like galaxies \citep{Daddi2005, Damjanov2009}.
We further extend this picture, finding a bimodal distribution
of bulges in terms of their formation redshift.
Indeed, these spheroids present an extended stellar disk 
component at the time of observation. 
Until today, great effort was put in studying the quiescent 
population at high redshift selecting it by means of the galaxy color.
Despite the UVJ selection is proven efficient in selecting genuinely quiescent 
galaxies, it could miss a significant fraction of mass hidden within spheroids in
luminous star-forming disk galaxies \citep{Schreiber2018}.

\subsection{Spheroids evolve in two modes}

\begin{figure}[t!]
\centering
\includegraphics[scale=0.30, trim=0cm 0cm 0cm 0cm , clip=true]{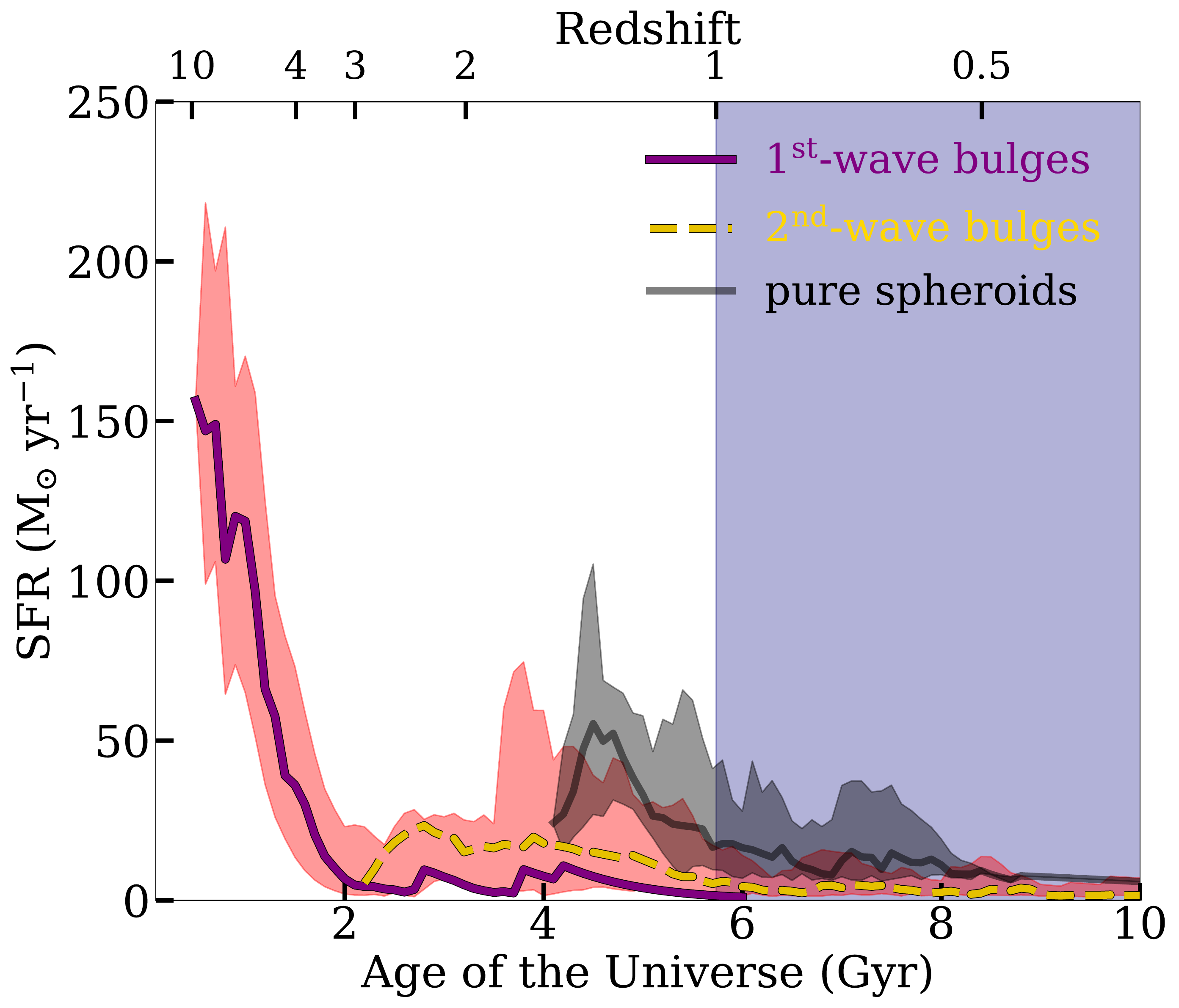}
\caption{Averaged SFHs of the 1$^{\rm st}$-wave bulges (purple solid line),
2$^{\rm nd}$-wave bulges (orange dashed line), and pure spheroids (black solid line).
The red and black shaded curves represent the corresponding 16th-84th percentile interval.
The blue shaded area indicates the redshift studied in this work.}
\label{fig:average_SFH}
\end{figure}

The question which arises is if the bimodal distribution in the ages of the 
spheroidal population is due to a different channel of evolution.
Looking at Fig.~\ref{fig:figure_8}, our analysis seems to suggest that 
between redshift $2 \lesssim z \lesssim 4$ the processes responsible for assembling the 
more massive spheroids were less efficient,
even though it remains difficult to state if it is a reflection 
of different pathways and transitional epochs in galaxies evolution.
In order to quantify the formation timescale of the spheroids in our sample,
we show in Fig.~\ref{fig:figure_11} their $\tau$ distributions.
The build up of the spheroidal component not only comes in two waves, 
but it seems to be characterized by two different modes of formation: a fast and a slow one.
The majority of our systems form in short timescales, having a median $\tau = 203_{-4}^{+160}$~Myr.
However, 1$^{\rm st}$-wave bulges have average timescale $\tau = 232 \pm 16$~Myr,
while 2$^{\rm nd}$-wave bulges present $\tau = 716  \pm 203$~Myr.
The scatter of the $\tau$ distribution is $\sim90$~Myr for 1$^{\rm st}$-wave bulges and $\sim1.6$~Gyr for 2$^{\rm nd}$-wave ones.
Among 2$^{\rm nd}$-wave bulges, 15 out of 61 (25\%) have $\tau>500$~Myr
and 31\% have $\tau>300$~Myr. On the other hand, 
97\% (29 our of 30) of 1$^{\rm st}$-wave bulges have $\tau<500$~Myr and 90\% have $\tau<300$~Myr.
Furthermore, we present in Fig.~\ref{fig:average_SFH} the averaged 
SFHs separately for bulges and pure spheroids.
The fast mode is mostly the only channel of evolution at earlier cosmic epochs ($z \gtrsim 5$).
On the other hand, the slow mode starts to be relevant only in the last $\sim10$~Gyr ($z \lesssim 2-3$).
This picture is consistent with an increasing number of results,
where the fast mode dominates at redshift $z\gtrsim2.5$
but the two channels have to be in place simultaneously and 
with similar quenching rates at $z\sim2$ \citep{Schawinski2014, Wild2016, Belli2019}. 

\begin{figure*}
\centering
\includegraphics[scale=0.52]{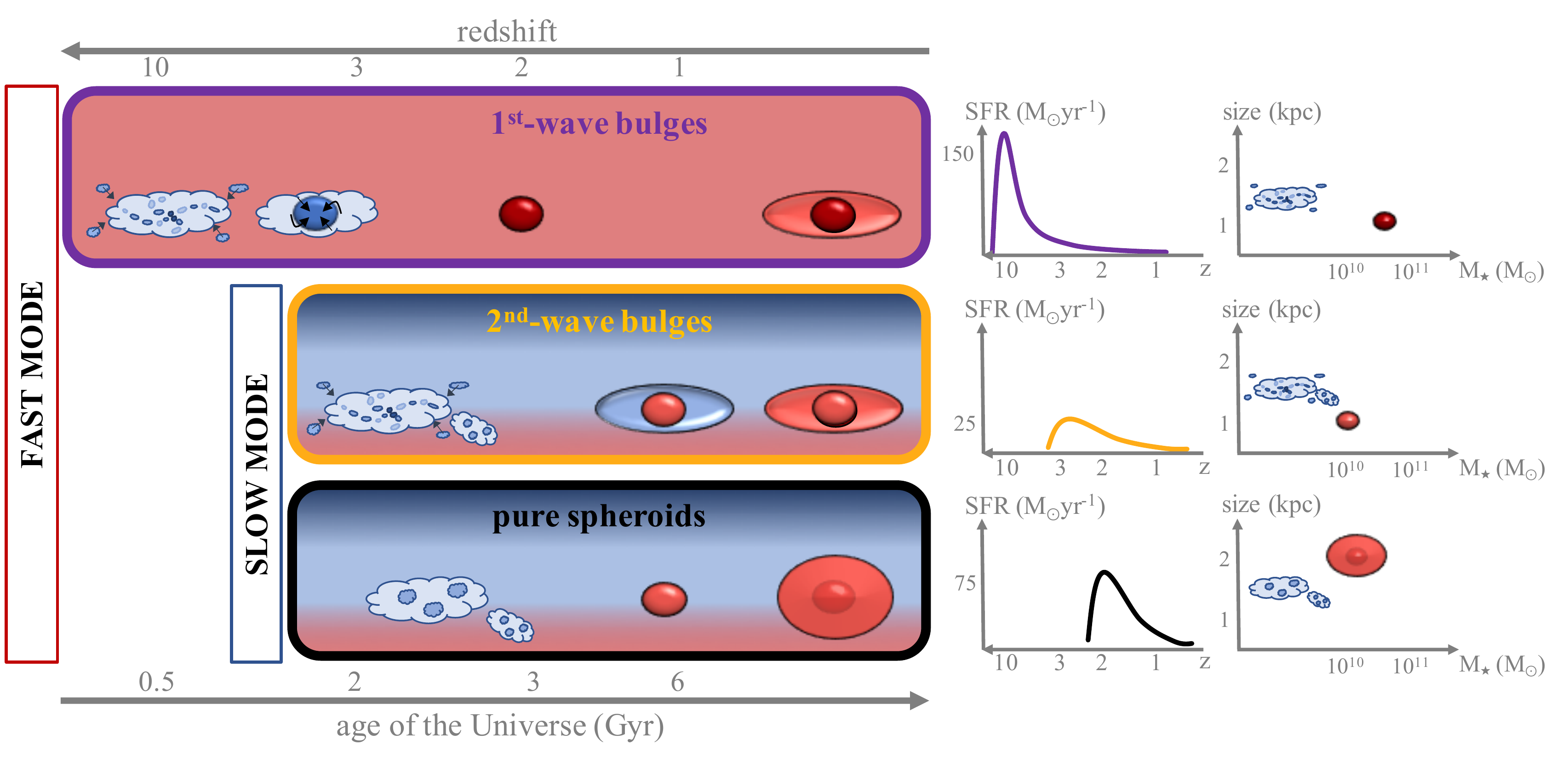}
\caption{Illustration of the proposed formation scenario for the 
spheroidal component of massive galaxies at redshift $0.14<z \leq1$.
\label{fig:cartoon}}
\end{figure*}

\begin{deluxetable*}{ccccccc}
\tablecaption{Median physical properties of the spheroidal population at redshift \mbox{$0.14 < z \leq 1$}. \label{tab:table_2}}
\tablehead{
\colhead{Type} & \colhead{$\log(M_{\star}$)} & \colhead{$\bar{t}_{M}$} & 
\colhead{$z_{\rm form}$} & \colhead{$\tau$} & \colhead{$R_{\rm e}$} & \colhead{$\log(\Sigma_{1.5})$} \\
\colhead{} & \colhead{(M$_{\odot}$)} & \colhead{(Gyr)} & 
\colhead{} &\colhead{(Myr)} & \colhead{(kpc)} & \colhead{(M$_{\odot}$ kpc$^{-1.5}$)} 
}
\decimalcolnumbers
\startdata
1$^{\rm st}$-wave bulges		&	$10.8^{+0.2}_{-0.3}$		&	$6.5^{+1.5}_{-1.4}$	&	$6.2^{+1.5}_{-1.7}$	&	$203^{+15}_{-3}$	&	$1.3^{+0.8}_{-0.6}$	&	$10.6^{+0.4}_{-0.4}$	\\
2$^{\rm nd}$-wave bulges		&	$10.3^{+0.4}_{-0.5}$		&	$1.7^{+2.0}_{-0.8}$	&	$1.3^{+0.6}_{-0.6}$	&	$213^{+790}_{-13}$	&	$1.0^{+0.8}_{-0.4}$	&	$10.2^{+0.5}_{-0.4}$	\\
pure spheroids				&	$10.5^{+0.3}_{-0.2}$ 	&	$1.3^{+1.2}_{-0.2}$	&	$1.1^{+0.3}_{-0.3}$ 	&	$201^{+114}_{-3}$	&	$1.9^{+1.4}_{-0.6}$	&	$10.1^{+0.2}_{-0.3}$	\\
\enddata
\tablecomments{(1) Spheroidal type; (2) Stellar mass of each component; (3) Mass-weighted age; (4) Redshift of formation;
(5) Timescale of exponentially declined SFH; (6) Effective radius; (7) Mass surface density.}
\end{deluxetable*}

\subsection{The old population of compact bulges}

Another piece of information is provided by the mass-size evolution of our spheroidal population.
Comparing systems at different formation redshift (see Fig.~\ref{fig:figure_9}),
we find that older bulges are more compact than younger spheroids (both bulges and pure spheroids).
This finding agrees with the fact that not only two waves of spheroids formation seems to be in place,
but fast-mode spheroids appear more compact than slow-mode ones \citep[see also][]{Belli2019}.
On the other hand, 1$^{\rm st}$ and 2$^{\rm nd}$-wave bulges have similar sizes of
$R_{\rm e} = 1.3^{+0.8}_{-0.6}$~kpc and $R_{\rm e} = 1.0^{+0.8}_{-0.4}$~kpc, respectively.
Despite being a result already observed in \citet{Whitaker2012} dividing
the red sequence of their quiescent galaxies into blue and red halves,
it is worth noting that this lack of evolution
is seen here for the first time in the bulge component.

As stated before, our results provide the evidence of a population of bulges
which formed by compaction in the early Universe. They are fast-track spheroids,
which present similar sizes but higher masses compared to their younger counterparts. 
In this context, we favor a scenario in which older spheroids in our sample 
where formed by violent disk instabilities and clumps migration, rather than mergers.
The higher abundance of cold gas available in the early Universe 
could provide the ideal conditions to rapidly generate very efficient starbursts,
which result in very compact and massive systems.
Cosmological simulations presented by \citet{Zolotov2015} show that their most
massive galaxies start their compaction phase at redshift $z\gtrsim4$ and successfully quench by redshift $z\sim2$.
Furthermore, recent results from cosmological simulations presented in \citet{Ceverino2018}
show that high $SFRs$ are driven by high gas accretion rates and the successive compaction of the stellar systems.
Indeed, in the case the compaction event induces an intense star formation burst 
at $z=10$ with maximum specific $SFR \sim 20$ Gyr$^{-1}$,
this translates to $\sim SFR=200$ M$_{\odot}$ yr$^{-1}$ for $10^{10}$ 
M$_{\odot}$ at $z=10$, mimicking the picture we are providing in Fig.~\ref{fig:average_SFH}.

The compaction phase of galactic disks into spheroids naturally leads to quench the 
systems stabilizing the violent disk instabilities. The direct result of building this 
mass concentration is rapidly increasing the angular velocity of the gas 
\citep[morphological quenching;][]{Martig2009}, diminishing the surface
density of the cold gas component \citep[i.e., star-formation and outflows;][]{Forbes2014},
and increasing the gas radial velocity dispersion \citep[i.e., stellar or AGN feedback;][]{Krumholz2013}.
The consequent suppression of the star formation is followed by a 
gradual growth and expansion into a larger spheroidal galaxy which could develop 
an extended stellar disk surrounding the spheroid due to late and slower gas accretion.
Indeed, at the time of observation, the bulge population presents an extended stellar disk,
which fits the more efficient redistribution of the gas 
driven by violent disk instabilities rather than mergers,
which could lead to a disruption of the disky kinematics, diminishing
more abruptly the angular momentum of the forming system.

\subsection{The second dominant wave of spheroids}

The totality of pure spheroids in our sample are formed at the epoch of the star-formation 
activity peak \citep{Lilly1996, Madau1996, Madau2014}.
Likewise, more than half of bulges were also formed in the last 10 Gyr of the Universe life.
This second (dominant) wave, which comprises a younger population of spheroids, is more challenging to interpret.
While the majority of spheroids are still formed in a fast mode, a slower mode starts to take place.
Again, there is a subpopulation of bulges which is more compact and formed in shorter timescales,
pointing towards a formation by compaction at delayed times respect to the older population.
On the other hand, there are pure spheroids and bulges which are on average less compact
and formed with a variety of timescales.
It is difficult to discriminate any scenario, and probably different ones start to contribute
at some level in shaping the properties of those spheroids.

Finally, we find that bulges are on average more compact than pure spheroids.
This made us question the true nature of these supposedly pure spheroids: are they 
hosting an inner (older) component which belongs to the same family of older bulges?
Theoretical arguments favor minor mergers to be responsible of their size growth, 
since it would be the physical processes allowing to efficiently increase the size of the spheroid 
compared to the growth of its stellar mass \citep{Naab2009, Hopkins2010}.
After the compaction phase, the high-redshift spheroid is likely to experience further accretion,
becoming more star-dominated as it evolves \citep{Oser2010, Porter2014}.
This wet and/or dry growth could be substantial, leading the spheroid to acquire 
an extended stellar envelope and making it appear like a typically observed 
elliptical galaxy in the local Universe \citep{vanDokkum2014, Buitrago2017}.

We qualitatively illustrate in Fig.~\ref{fig:cartoon} the proposed scenario for the formation and evolution 
of bulges and pure spheroids in massive galaxies at $0.14<z \leq1$. 
Moreover, we summarize the main properties of the different spheroids in Table~\ref{tab:table_2}.
The bulges in $z\leq1$ massive disk galaxies form in two waves. 
A first wave of bulges builds up fast in the early Universe ($z>3$) 
through very dissipative processes such as violent disk instabilities. 
They undertake a compaction phase, probably evolving to the 
well-known $z\sim2$ red nuggets, and then acquire a disk. 
A second wave of spheroids forms at $z=1-2$, some of them accrete 
new material and acquire a disk by $z\sim0.5$, becoming a bulge, 
some of them reach our sample as a pure spheroid. 
The first wave is characterized by short formation timescales and large peak SFR. 
A relatively slower mode of formation starts to be relevant for the 2$^{\rm nd}$-wave bulges 
and pure spheroids, presenting longer timescales and smaller peak SFRs. 
As mentioned earlier, 1$^{\rm st}$-wave bulges are characterized by their compactness: 
they present similar sizes but are more massive than 2$^{\rm nd}$-wave bulges. 
Pure spheroids are larger than bulges and present similar masses, i.e., they are not as compact.

%%%%%%%%%%%%%%%%%%%%%%%%%%%%%%%%%%%%%%%%%%%%%%%%%%%%%%%%%%%
%%%%%%%%%%%%%%%%%%%%%%%%%%%%%%%%%%%%%%%%%%%%%%%%%%%%%%%%%%%
%% SECTION 6

\section{Conclusions} \label{sec:section_6}

In this work we investigate the assembly history and evolutionary pathways 
of the spheroidal structures within massive galaxies at redshift $z\leq1$.
For that purpose, we use a sample of massive galaxies selected by 
the SHARDS spectro-photometric survey in GOODS-N.
We decomposed spectro-photometrically the light of our galaxies
into two main stellar components (i.e., the central bulge and the outer extended disk), 
retrieving their separate SEDs with a resolution $R\sim50$. 
We fit the photometry for the spheroids to stellar population synthesis models,
characterizing the SFH of the 91 bulges and comparing their properties
with the 65 pure spheroids in our sample, all at redshift $0.14 < z \leq 1$.

By deriving the mass-weighted ages and the formation redshifts of spheroids,
we find that they form in two waves. We distinguish
a fraction of very old bulges, which are formed at redshift ($z_{\rm form} > 3$),
and a dominant population of spheroids (bulges and pure spheroids) formed at 
median redshift $z_{\rm form} = 1.2_{-0.4}^{+0.4}$.
Furthermore, spheroids not only form in two waves,
but they also form in two modes.
At higher redshift, a fast mode (timescale around 200~Myr) 
is driving the rapid evolution of those systems,
while at lower redshift a slower mode starts to became relevant 
(with timescales ranging from 200~Myr to 1~Gyr and longer).
Finally, the old population of bulges is more compact than the young
population of spheroids. 
We propose that the first wave of formation is characterized by a violent compaction phase,
which builds up distinctively dense spheroids in short timescales.

Considering the rarity of high-$z$ red nuggets surviving in the local Universe 
\citep{Trujillo2009, Tortora2016, Charbonnier2017, Buitrago2018},
our results have important implication for the evolution of this population.
The population of red nuggets, formed at high redshift through rapid compaction \citep{Damjanov2009, Dekel2009},
could possibly evolve in today's early type galaxies or settle in the center of a disk galaxy.
Recently, \citet{Costantin2020} suggested that disk galaxies hide the remnants of the 
compact and quiescent population observed at high redshift.
The results of this work reinforce the increasing 
evidence that the central regions of early-type galaxies
actually harbors the population of spheroids formed at high 
redshift \citep{MacArthur2003, Graham2014, Graham2015},
broadening this comprehensive and compelling picture 
to later Hubble types.

%%%%%%%%%%%%%%%%%%%% FACILITIES %%%%%%%%%%%%%%%%%

%\facilities{GTC(OSIRIS), HST(ACS, WFC3), CFHT(WIRCam)}

%%%%%%%%%%%%%%%%%%%% ACKNOWLEDGEMENTS %%%%%%%%%%

\acknowledgments

We would like to thank the anonymous referee for improving the content of the manuscript.
LC wishes to thank Cristina~Cabello for the support provided
while this project was devised and Michele~Perna for the useful discussion.
\\
LC acknowledges financial support from Comunidad de Madrid under 
Atracci\'on de Talento grant 2018-T2/TIC-11612.
LC and PGPG acknowledge support from
Spanish Ministerio de Ciencia, Innovaci\'on y Universidades through grant PGC2018-093499-B-I00.
JMA acknowledges support from the MCIU by the grant AYA2017-83204-P and the 
Programa Operativo FEDER Andaluc\'ia 2014-2020 in collaboration with the 
Andalucian Office for Economy and Knowledge.
DC is a Ramon-Cajal Researcher and is supported by the Ministerio de Ciencia, 
Innovaci\'{o}n y Universidades (MICIU/FEDER) under research grant PGC2018-094975-C21.
\\
This work has made use of the Rainbow Cosmological Surveys Database, 
which is operated by the Centro de Astrobiolog\'ia (CAB/INTA), 
partnered with the University of California Observatories at Santa Cruz (UCO/Lick,UCSC).

%%%%%%%%%%%%%%%%%%%% APPENDIX %%%%%%%%%%%%%%%%%

\clearpage

\appendix

\section{Spectro-photometric decoupling: robustness of the method \label{sec:appendix_A}}

In this Section we describe the robustness of our decomposition process 
and provide some more examples of the spectro-photometric decoupling performances.

\begin{figure}[t]
\centering
\includegraphics[scale=0.31]{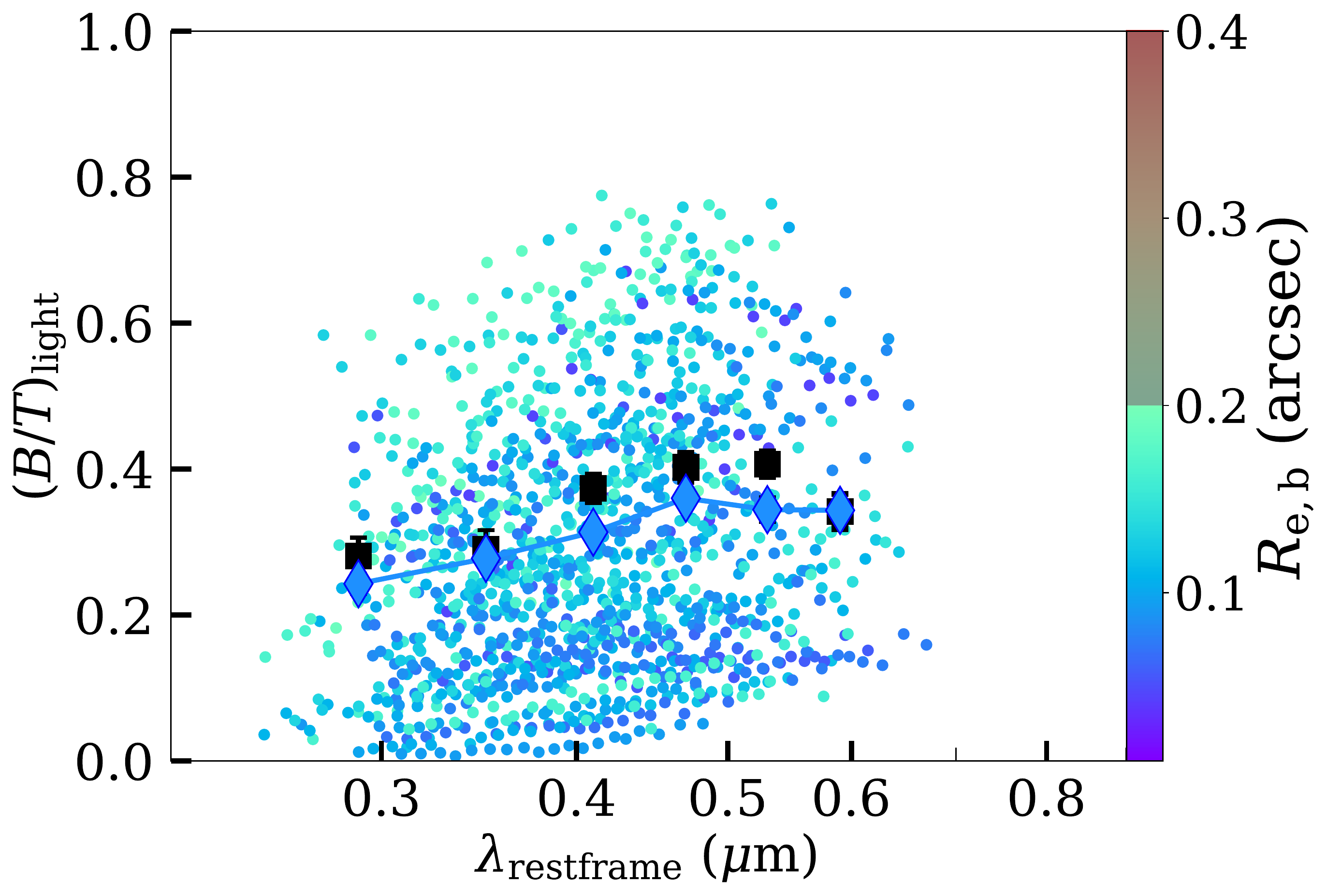}
\hspace{0.4cm}
\includegraphics[scale=0.31]{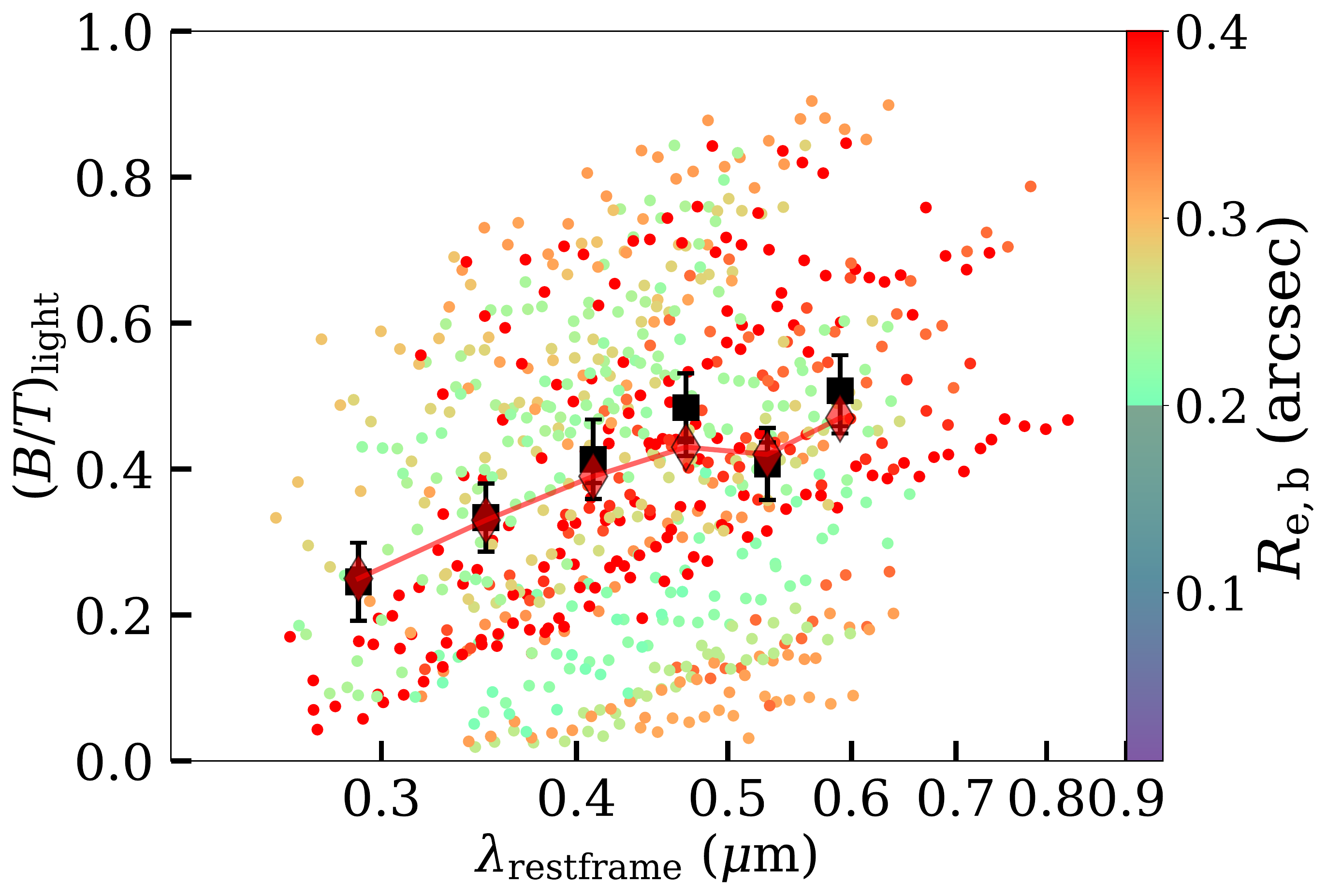}
\caption{Luminosity-weighted $B/T$ as a function of wavelength (dots)
for galaxies with $R_{\rm e,b} < 0.2$ arcsec (left panel) and  $R_{\rm e,b} > 0.2$ arcsec (right panel).
Dots are color-coded according to the bulge effective radius. 
Diamonds represent the mean values measured from our SHARDS photometry,
while squares stand for mean values retrieved from \citet{Dimauro2018}.
Error bars stand for the standard error of the mean.}
\label{fig:appendix_A0}
\end{figure}

As described in Sect.~\ref{sec:section_3-1}, we combined the HST and SHARDS data 
to get SEDs for bulges and pure spheroids with a spectral resolution $R\sim50$. 
In this sense, we did not fit two components blindly and independently to the SHARDS data, 
but we use the decomposition of the HST images as priors.
We fit the characteristic surface brightness levels of bulges and disks in SHARDS bands,
keeping their structural parameters (i.e., S\'ersic index, size, and shape) 
frozen from the two-dimensional decomposition of the galaxy light in HST filters.
Since only the relative intensity of the bulge and disk component  
is allowed to vary in the decomposition of SHARDS images, 
we compare in Fig.~\ref{fig:appendix_A0} the trend of $(B/T)_{\rm HST}$ and $(B/T)_{\rm SHARDS}$,
dividing the galaxies according to the size of their bulges.
The sensibility of our decomposition process is presented
in Tables~\ref{tab:appendix_A1-1} and \ref{tab:appendix_A1-2}.
We find that the average relative error between $(B/T)_{\rm HST}$ and $(B/T)_{\rm SHARDS}$
is less than 16\% for galaxies hosting small bulges ($R_{\rm e,b} < 0.2$~arcsec) and
it is less than 11\% for galaxies hosting large bulges ($R_{\rm e,b} > 0.2$~arcsec).

\begin{figure*}[h!]
\centering
\includegraphics[scale=0.9, trim=0cm 0cm 0cm 0cm, clip=true]{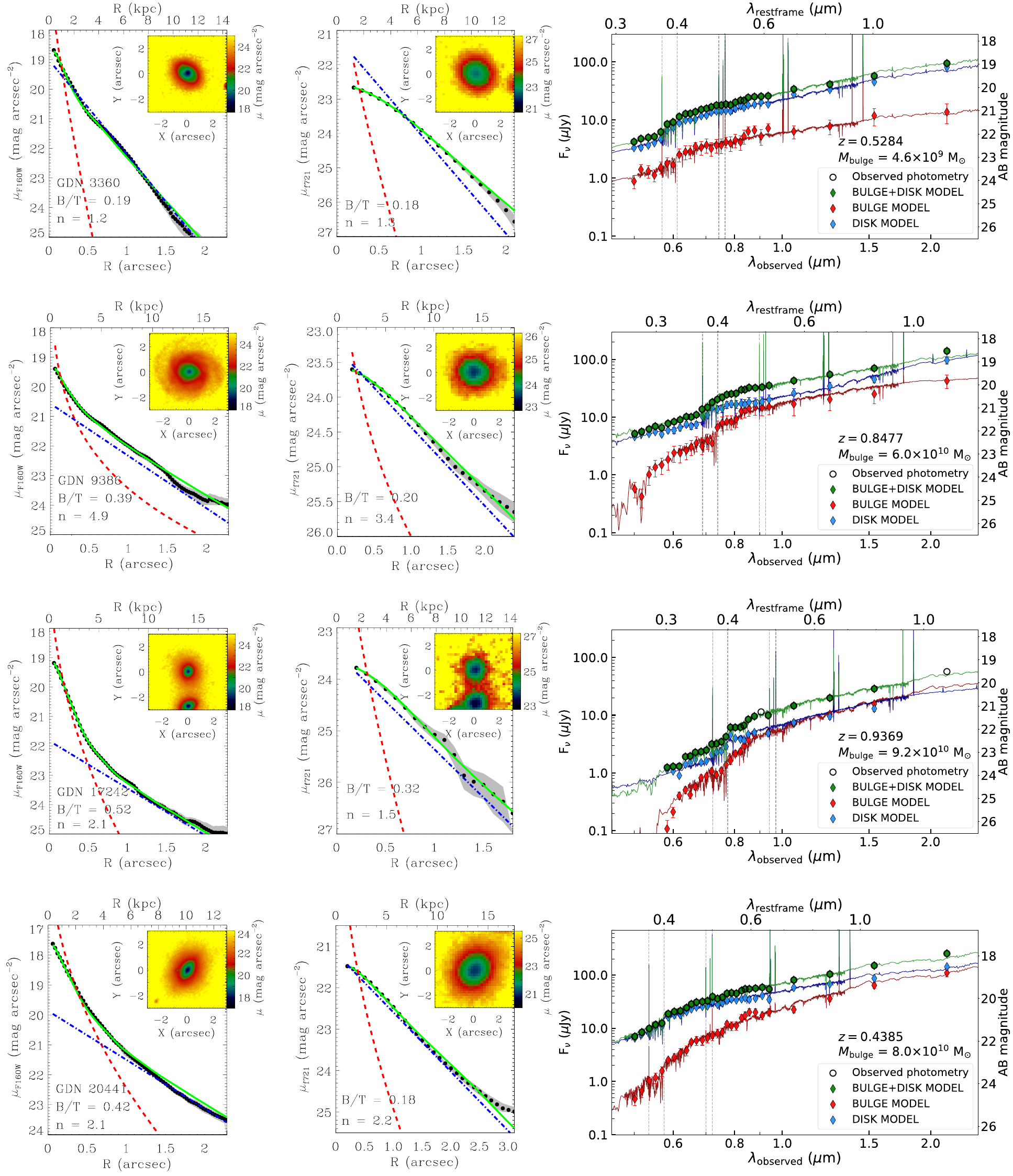}
\caption{Two-dimensional photometric decomposition on HST (left panels) and SHARDS
(middle panels) images, similarly to those proposed in Figs.~\ref{fig:figure_3} and~\ref{fig:figure_4}. 
The SEDs and the best fit stellar models of the galaxy, bulge, and disk are also presented (right panels),
similarly to those proposed in Fig.~\ref{fig:figure_6}. 
From top to bottom, the galaxies GDN~3360, GDN~9386,  GDN~17242, and GDN~20441 are shown as an example.
\label{fig:appendix_A1}}
\end{figure*}

We randomly pick four galaxies out of our sample 
(i.e., GDN~3360, GDN~9386, GDN~17242, and GDN~20441) in order to demonstrate 
how the two-dimensional photometric decomposition performed on HST images transfers 
to SHARDS data, as detailed in Sect.~\ref{sec:section_3-1}.
We show in Fig.~\ref{fig:appendix_A1} how the synergy between the two data set allows us to fit the individual SED of bulges 
and disks (right panels), characterizing their SFHs, as detailed in Sect.~\ref{sec:section_3-2}. 

\section{Mass-weighted ages: degeneracies \label{sec:appendix_B}}

\begin{figure*}[h!]
\centering
\includegraphics[scale=0.4, trim=0.5cm 0cm 0.5cm 1cm , clip=true]{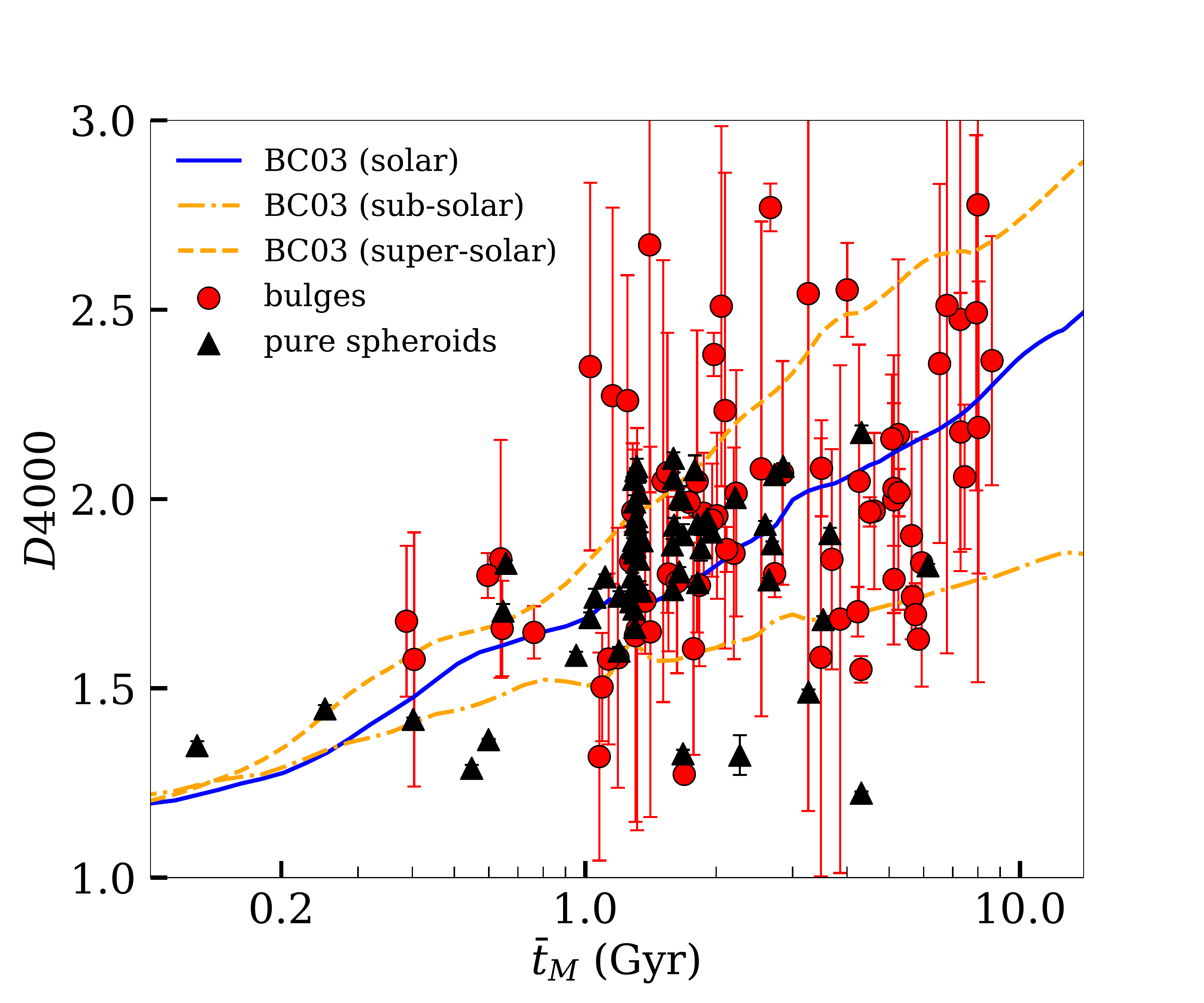}
\caption{Measured $D4000$ values as a function of mass-weighted ages
for bulges (red dots) and pure spheroids (black triangles). 
The solid blue line shows the evolution of $D4000$ \citet{Bruzual2003} models at solar metallicity, 
while the dashed and dashed-dotted orange lines correspond to super-solar and sub-solar 
metallicities, respectively.}
\label{fig:appendix_B0}
\end{figure*}

It is well known that stellar age, dust content, and metallicity
are strongly degenerate, introducing uncertainties in physical parameter estimates based on SED fitting.
However, key spectral features in the optical rest-frame spectra of galaxies (i.e., the 4000~\AA~break and the Balmer break)
better constrain the typical degeneracies which affect the SFH of each galaxy
\citep{Ferreras2012, Whitaker2011}. In particular,
since the break in the stellar continuum at 4000~\AA~is a good age indicator 
\citep{Bruzual1983, Kauffmann2003},
it has been successfully used to infer redshifts and ages 
of stellar populations in massive galaxies at high-$z$ \citep{Saracco2005}, 
and can also be successfully measured using medium-band photometry \citep{Kriek2011}.
It is worth noting that, since we are dealing with estimations from photometric data alone,
we decide to use the definition in \citet{Bruzual1983} instead of the narrow index $D_{n}4000$ \citep{Balogh1999},
in order to reduce the uncertainties in the measurement.
The SHARDS data provide a combination of depth and spectral resolution
which allows us to detect such feature not only in the galaxy integrated SED but even in the bulge one
(see Fig.~\ref{fig:figure_6}), providing an independent validation of the its age.
Thus, to assess the robustness of the mass-weighted ages retrieved from
the SED fitting proposed in Sect.~\ref{sec:section_3-2} and~\ref{sec:section_4-1}, 
we show in Fig.~\ref{fig:appendix_B0} our measurements of $D4000$ for the sample 
bulges and pure spheroids, comparing them with the trend expected from \citet{Bruzual2003} stellar population models.

\begin{figure*}[h!]
\centering
\includegraphics[scale=0.3, trim=0cm 0cm 0cm 0cm, clip=true]{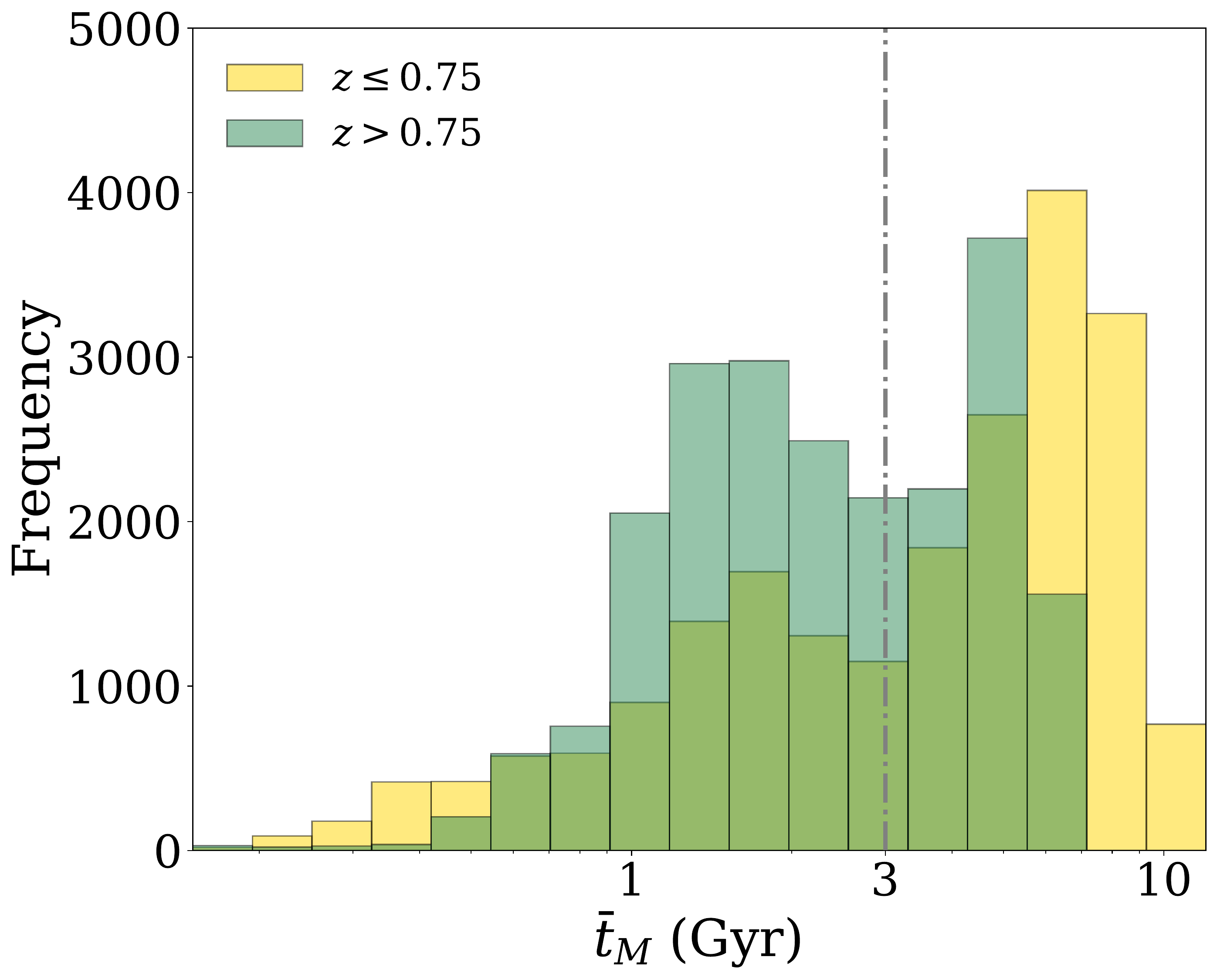}
\caption{Distribution of mass-weighted age considering
the 500 MC simulations for each bulge in our sample.
The yellow histogram represents bulges in galaxies at redshift $z \leq 0.75$, while the green 
histogram stands for bulges in galaxies at redshift $z > 0.75$.
\label{fig:appendix_B1}}
\end{figure*}

\begin{figure*}[h!]
\centering
\includegraphics[scale=0.4, trim=0cm 0cm 0cm 0cm, clip=true]{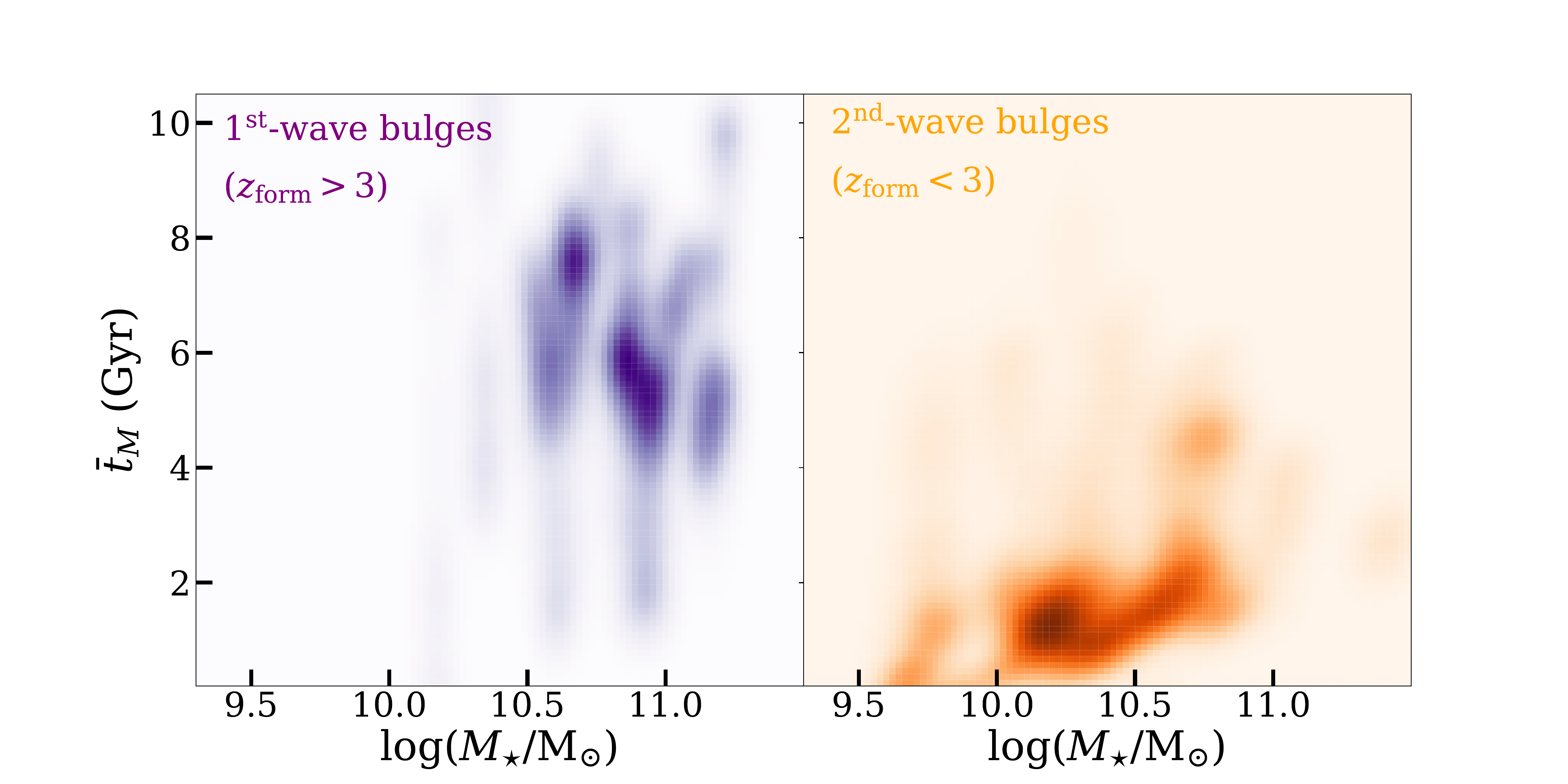}
\caption{Density distribution of the mass-weighted ages of our bulges retrieved 
from the MC simulations described in Sect.~\ref{sec:section_3-2}. Bulges are separated 
between 1$^{\rm st}$-wave (purple distribution in the left panel) and 2$^{\rm nd}$-wave ones 
(orange distribution in the right panel).
Darker colors stand for higher density.
\label{fig:appendix_B2}}
\end{figure*}

In Fig.~\ref{fig:appendix_B1} we show the distribution of mass-weighted ages
of the bulges in our sample, dividing them into those observed at redshift 
$z \leq 0.75$ (46 out of 91) and those at redshift $z>0.75$ (45 out of 91).
We see that the trends presented in Fig.~\ref{fig:figure_7} (and Fig.~\ref{fig:figure_8})
hold even when we separate bulges in lower and higher redshift ones. 
This allows us to rule out that the bimodality of bulge ages
is biased because of their redshift distribution and/or large scale structure (see Fig.~\ref{fig:figure_1}).
Additionally, to further quantify that the age bimodality reported
in Sect.~\ref{sec:section_4-1} is not driven by degeneracies in 
the cluster of solutions resulting from the SED fitting, we show in Fig.~\ref{fig:appendix_B2} the 
distribution of the 500 MC simulations performed for every bulge in our sample.
In particular, we separate 1$^{\rm st}$-wave bulges ($z_{\rm form} > 3$; left panel)
from 2$^{\rm nd}$-wave ones ($z_{\rm form} < 3$; right panel).
We show how the age bimodality holds independently of choosing the best cluster solution,
implying a physical separation between the two waves of bulges.

\begin{deluxetable*}{cccccc}
\tablecaption{Comparison of $B/T$ from SHARDS and HST photometry for galaxies with $R_{\rm e,b} < 0.2$~arcsec. \label{tab:appendix_A1-1}}
\tablehead{
\colhead{$\lambda_{\rm rest}$} & 
\colhead{$(B/T)_{\rm HST}$} &
\colhead{$\sigma_{\rm HST}$} &
\colhead{$(B/T)_{\rm SHARDS}$} &
\colhead{$\sigma_{\rm SHARDS}$} &
\colhead{Relative difference} \\
\colhead{(nm)} & 
\colhead{} &
\colhead{} &
\colhead{} &
\colhead{} &
\colhead{}}
\decimalcolnumbers
\startdata
$0.26 - 0.32$		&	$0.28 \pm 0.02$	&	$0.16$		&	$0.24 \pm 0.01$	&	$0.14$		&	$14\%$		\\
$0.32 - 0.38$		&	$0.29 \pm 0.03$	&	$0.12$		&	$0.28 \pm 0.01$	&	$0.15$		&	$4\%	$		\\
$0.38 - 0.44$		&	$0.37 \pm 0.02$	&	$0.13$		&	$0.31 \pm 0.01$	&	$0.17$		&	$16\%$		\\
$0.44 - 0.50$		&	$0.40 \pm 0.02$	&	$0.13$		&	$0.36 \pm 0.01$	&	$0.19$		&	$11\%$		\\
$0.50 - 0.56$		&	$0.41 \pm 0.02$	&	$0.10$		&	$0.34 \pm 0.02$	&	$0.19$		&	$15\%$		\\
$0.56 - 0.62$		&	$0.34 \pm 0.03$	&	$0.12$		&	$0.34 \pm 0.02$	&	$0.16$		&	$<1\%$		\\
\enddata
\tablecomments{(1) Rest-frame wavelength range; 
(2) Mean bulge-over-total luminosity ratio and standard error \citep[HST;][]{Dimauro2018};
(3) Standard deviation;
(4) Mean bulge-over-total luminosity ratio and standard error (SHARDS); 
(5) Standard deviation;
(6) Relative difference between $(B/T)_{\rm HST}$ and $(B/T)_{\rm SHARDS}$.}
\end{deluxetable*}

\begin{deluxetable*}{cccccc}
\tablecaption{As Table~\ref{tab:appendix_A1-1}, but for galaxies with $R_{\rm e,b} > 0.2$~arcsec. \label{tab:appendix_A1-2}}
\tablehead{
\colhead{$\lambda_{\rm rest}$} & 
\colhead{$(B/T)_{\rm HST}$} &
\colhead{$\sigma_{\rm HST}$} &
\colhead{$(B/T)_{\rm SHARDS}$} &
\colhead{$\sigma_{\rm SHARDS}$} &
\colhead{Relative difference} \\
\colhead{(nm)} & 
\colhead{} &
\colhead{} &
\colhead{} &
\colhead{} &
\colhead{}}
\decimalcolnumbers
\startdata
$0.26 - 0.32$		&	$0.25 \pm 0.05$	&	0.24		&	$0.25 \pm 0.02$	&	0.16		&	4\%	 			\\
$0.32 - 0.38$		&	$0.33 \pm 0.05$	&	0.19		&	$0.33 \pm 0.02$	&	0.19		&	$<1$\% 	 		\\
$0.38 - 0.44$		&	$0.41 \pm 0.05$	&	0.24		&	$0.39 \pm 0.01$	&	0.20		&	6\%		 		\\
$0.44 - 0.50$		&	$0.48 \pm 0.05$	&	0.22		&	$0.43 \pm 0.02$	&	0.22		&	11\%		 	 	\\
$0.50 - 0.56$		&	$0.41 \pm 0.05$	&	0.21		&	$0.42 \pm 0.02$	&	0.22		&	4\%		 	 	\\
$0.56 - 0.62$		&	$0.51 \pm 0.05$	&	0.22		&	$0.47 \pm 0.02$	&	0.19		&	7\%			 	\\
\enddata
\tablecomments{(1) Rest-frame wavelength range; 
(2) Mean bulge-over-total luminosity ratio and standard error \citep[HST;][]{Dimauro2018};
(3) Standard deviation;
(4) Mean bulge-over-total luminosity ratio and standard error (SHARDS); 
(5) Standard deviation;
(6) Relative difference between $(B/T)_{\rm HST}$ and $(B/T)_{\rm SHARDS}$.}
\end{deluxetable*}

\clearpage

%%%%%%%%%%%%%%%%%%%% REFERENCES %%%%%%%%%%%%%%%%%

\bibliography{../../Costantin}{}
\bibliographystyle{aasjournal}

\end{document}